\documentclass[aps,
prd,
twocolumn,
10pt,
preprint,
tightenlines,
superscriptaddress,
nofootinbib,
showpacs
]{revtex4}
\usepackage{amssymb,latexsym}
\usepackage{amsmath,amsbsy,bbm}
\usepackage{epsfig,bm}
\usepackage{graphicx,comment}
\DeclareMathOperator{\Tr}{Tr}
\DeclareMathOperator{\diag}{diag}
\DeclareMathOperator{\Erfc}{Erfc}

\begin{document}

\def\a{{\alpha}}
\def\b{{\beta}}
\def\d{{\delta}}
\def\D{{\Delta}}
\def\X{{\Xi}}
\def\e{{\varepsilon}}
\def\g{{\gamma}}
\def\G{{\Gamma}}
\def\k{{\kappa}}
\def\l{{\lambda}}
\def\L{{\Lambda}}
\def\m{{\mu}}
\def\n{{\nu}}
\def\o{{\omega}}
\def\O{{\Omega}}
\def\S{{\Sigma}}
\def\s{{\sigma}}
\def\th{{\theta}}

\def\cF{{\mathcal F}}
\def\cS{{\mathcal S}}
\def\cC{{\mathcal C}}
\def\cB{{\mathcal B}}
\def\cT{{\mathcal T}}
\def\cQ{{\mathcal Q}}
\def\cL{{\mathcal L}}
\def\cO{{\mathcal O}}
\def\cA{{\mathcal A}}
\def\cQ{{\mathcal Q}}
\def\cR{{\mathcal R}}
\def\cH{{\mathcal H}}
\def\cW{{\mathcal W}}
\def\cM{{\mathcal M}}
\def\cD{{\mathcal D}}
\def\cN{{\mathcal N}}
\def\cP{{\mathcal P}}
\def\cK{{\mathcal K}}

\def\ol#1{{\overline{#1}}}

\def\sumint{\sum \hskip-1.35em \int_{ \hskip-0.25em \underset{y,z}{\phantom{a}}} \,}

\def\eqref#1{{(\ref{#1})}}

\preprint{RBRC-1061}

\title{Neutron in a Strong Magnetic Field: Finite Volume Effects}

\author{Brian~C.~Tiburzi}
\email[]{btiburzi@ccny.cuny.edu}

\affiliation{Department of Physics, The City College of New York, New York, NY 10031, USA}
\affiliation{Graduate School and University Center, The City University of New York, New York, NY 10016, USA}
\affiliation{RIKEN BNL Research Center, Brookhaven National Laboratory, Upton, NY 11973, USA}

\date{\today}


\begin{abstract}
We investigate the neutron's response to magnetic fields on a torus with the aid of chiral perturbation theory, 
and expose effects from non-vanishing holonomies. 
The determination of such effects necessitates non-perturbative treatment of the magnetic field; 
and, 
to this end, 
a strong-field power counting is employed.
Using a novel coordinate-space method, 
we find the neutron propagates in a coordinate-dependent effective potential that we obtain by integrating out charged pions winding around the torus.  
Knowledge of these finite volume effects will aid in the extraction of neutron properties from lattice QCD computations in external magnetic fields. 
In particular, 
we obtain finite volume corrections to the neutron magnetic moment and magnetic polarizability.
These quantities have not been computed correctly in the literature.  
In addition to effects from non-vanishing holonomies, 
finite volume corrections depend on the magnetic flux quantum through an 
Aharonov-Bohm effect. 
We make a number of observations that demonstrate the importance of non-perturbative effects from strong magnetic fields currently employed in lattice QCD calculations. 
These observations concern neutron physics in both finite and infinite volume. 
\end{abstract}

\maketitle

\section{Introduction} %

Electromagnetic interactions of quarks are simple at the level of the action, 
however, 
the confining dynamics of quarks in QCD lead to electromagnetic properties of non-perturbative bound-state hadrons. 
While these properties can be parameterized in terms of a few low-energy hadronic parameters, 
determining their values from QCD provides a window to hadron structure in terms of quark degrees of freedom. 
Additionally the QCD response to external electromagnetic fields provides a lever-arm with which to study the modification of hadronic structure.  
For example, 
the energy levels of a neutron in a weak magnetic field depend on the neutron's magnetic moment and magnetic polarizability.
The former hadronic parameter provides a glimpse at the current distribution within the neutron, 
while the latter parameter encompasses the response of this current distribution to an external magnetic field. 
Lattice QCD computations in classical electromagnetic fields constitute a fruitful method to determine 
hadronic properties,
especially polarizabilities, 
see~\cite{%
Martinelli:1982cb,
Bernard:1982yu,
Fiebig:1988en,
Lee:2005ds,
Christensen:2004ca,
Lee:2005dq,
Shintani:2006xr,
Engelhardt:2007ub,
Shintani:2008nt,
Aubin:2008qp,
Detmold:2009dx,
Alexandru:2009id,
Detmold:2010ts,
Alexandru:2010dx,
Freeman:2012cy,
Primer:2013pva,
Lujan:2013qua,
Lujan:2014kia%
}.

Lattice gauge theory techniques allow one to determine the neutron magnetic moment and magnetic polarizability by studying neutron correlation functions in external magnetic fields. 
Because the notion of a magnetic moment, 
or a dipole polarizability, 
is intimately linked with rotational invariance, 
lattice calculations of these quantities are subject to non-trivial finite volume effects.
Indeed the imposition of a uniform magnetic field on a torus is already constrained by the finite volume%
~\cite{'tHooft:1979uj,Smit:1986fn,Damgaard:1988hh}.
The magnetic flux quantization for a torus leads to two notable effects. 
First is the restriction to magnetic fields that are not considerably small compared to hadronic scales.  
As a result, 
lattice computations are often probing hadronic response to strong magnetic fields. 
Second is the accompanying gauge holonomy, 
which is an artifact of the finite volume. 
Both of these effects complicate external field spectroscopy.%
\footnote{
Dirichlet boundary conditions for quarks have been proposed to overcome the difficulties associated with uniform electromagnetic fields on a torus. 
This choice introduces finite-size effects that are unfortunately rather complicated to address, 
see%
~\cite{Tiburzi:2013vza}
for a discussion of one such effect.
}

To address rigorously the response of the neutron in a strong magnetic field, 
we use chiral perturbation theory and combine three crucial ingredients for the first time. 
We use the heavy-nucleon chiral expansion to address nucleon-pion dynamics. 
Secondly the magnetic field is treated non-perturbatively by summing charged pion couplings to the external field with the aid of Schwinger's proper-time trick%
~\cite{Schwinger:1951nm}. 
In this way, 
effects of charged pion Landau levels are included in our computation. 
Thirdly the computation is performed in finite volume, 
and properly includes topological effects from charged pions winding around the torus, 
and the flux quantization for magnetic fields.%
\footnote{
Recently an analysis of finite volume corrections to the neutron magnetic polarizability appeared%
~\cite{Hall:2013dva}, 
however, 
their approach does not account for effects addressed in our work. 
} 
A few highlights of the various results obtained are as follows:

\begin{itemize}

\item
We obtain the infinite volume spin-independent energy shift, and Zeeman splitting of the neutron in strong magnetic fields, 
Eqs.~\eqref{eq:E0nonP} and \eqref{eq:E1nonP}.
These results are required to address the breakdown of perturbative magnetic field expansions, 
which is necessitated by the size of current-day lattices.

\item
We obtain expressions for finite volume corrections to the neutron magnetic moment and magnetic polarizability by considering what we deem to be extreme limits, 
Eqs.~\eqref{eq:FVmoment} and \eqref{eq:betaFV}. 
These results differ from those in the literature, 
however, 
those calculations are known to be inconsistent by keeping the lattice size
$L$
finite in one part of the calculation, 
and 
$L = \infty$
in another. 
We caution lattice practitioners from using such formulae due to the extremes required for their applicability.

\item
We obtain finite volume corrections to the neutron magnetic moment and polarizability that can be used in practice for external field computations in lattice QCD, 
Eqs.~\eqref{eq:kappa3} and \eqref{eq:beta3}.
The simplicity of these results reflects that charged pion winding transverse to the magnetic field is suppressed due to Landau level confinement. 
At finite volume, 
we show non-perturbative magnetic field effects also remain relevant on current-day lattices.

\end{itemize}

The organization of our presentation is as follows. 
We begin by reviewing the treatment of uniform magnetic fields on a torus in 
Sec.~\ref{s:Gen}. 
Included in this treatment is the solution to the magnetic periodic Green's function for charged scalars, 
and that for a charged heavy fermion in the static approximation. 
The one-loop computation of the neutron effective action in a magnetic field is pursued in 
Sec.~\ref{s:One}. 
The computation uses heavy nucleon chiral perturbation theory with strong-field power counting in the so-called
$p$-regime, 
and details pertinent to the computation are reviewed. 
We employ a direct coordinate-space approach that is first exemplified by computing finite volume corrections to the neutron mass.  
By extending the computation to the neutron two-point function in an external magnetic field, 
expressions are obtained for the spin-dependent and spin-independent terms in the neutron effective potential. 
The effective potential is coordinate dependent, 
which directly reflects the non-vanishing holonomies of the gauge field.
Perturbative expansions in powers of the magnetic field are contrasted with non-perturbative results obtained in strong-field chiral perturbation theory. 
The final section, 
Sec.~\ref{s:Disc}, 
concerns assessing the effect of finite volume. 
In this section,  
we evaluate corrections to the neutron magnetic moment and magnetic polarizability. 
While the effects are complicated by the coordinate dependence of the effective potential,
we find that such corrections to the ground-state neutron energy are suppressed. 
At the end of this section, 
we conclude with a few directions for future work. 
Various technical details have been relegated to appendices. 
In Appendix~\ref{s:D}, 
we detail the inclusion of delta-pion intermediate states in our calculation. 
The main text discusses only proton-pion intermediate states for simplicity. 
One-dimensional integral formulae for finite volume effects are collected in 
Appendix~\ref{s:A}. 
A final appendix, 
Appendix~\ref{s:B}, 
demonstrates how finite volume corrections to the magnetic moment are related between two different extreme limits.

\section{Generalities} 
\label{s:Gen} %

To begin, 
we discuss the inclusion of a uniform magnetic field on a torus, 
and the implications for charged particle two-point correlation functions. 
Such generalities will be required to pursue chiral perturbation theory  computations at one-loop order for the neutron. 
We consider four-dimensional Euclidean spacetime,
with three directions compactified to form a torus.  
The length of each compact direction is taken to be 
$L$. 
The fourth direction will correspond to Euclidean time and is non-compact. 
Throughout we use Greek indices for quantities having four components, 
$x_\mu = (x_1, x_2, x_3, x_4)$, 
and Latin indices for quantities having three spatial components, 
$x_j$. 
We additionally employ arrow notation,
$\vec{x} = ( x_1, x_2, x_3, 0)$, 
for quantities with three spatial components, 
and hat notation, 
$\hat{x}_j$,
for unit spatial vectors. 
Bold symbols are reserved for quantities having the particular form 
$\bm{x} = (x_1, 0, x_3, x_4)$, 
i.e.~a quantity that appears similar to a four-vector but with the second component deleted. 
Lastly 
we will often append a subscript 
$\perp$
to denote the spatial directions perpendicular to the magnetic field.

\subsection{Magnetic Field}

In infinite volume, 
the vector potential 
$A_\mu(x) = ( - B x_2, 0, 0, 0)$
gives rise to a uniform magnetic field, 
$\vec{B} = B \hat{x}_3$. 
On a torus, 
however,
one needs to worry about boundary conditions. 
A particularly clear exposition of these issues is given in%
~\cite{AlHashimi:2008hr}.
While a uniform magnetic field is trivially periodic, 
the gauge potential 
$A_\mu (x)$
is not periodic, 
but is periodic only up to a gauge transformation, 
namely
\begin{equation}
A_\mu ( x + L \hat{x}_j ) = A_\mu (x) + \partial_\mu \Lambda_j (x)
\label{eq:nonP}
,\end{equation}
where the transformation function is given by 
$\Lambda_j (x) = ( 0, - B L x_1, 0)$. 
On a torus, 
moreover, 
the electric and magnetic fields are not the only gauge invariant quantities. 
There are additionally holonomies of the gauge field which can be expressed in terms of Wilson lines spanning the compact directions.
For our application, it is useful to consider a matter field of electric charge
$Q$, 
for which the Wilson lines have the form 
\begin{equation}
W_j (x)
= 
e^{i Q \int_0^L dx_j A_j(x)}
e^{ - i Q \Lambda_j(x)}
,\end{equation} 
where the Einstein summation convention has been suspended. 
For the linearly rising gauge potential,
\footnote{
One can also consider the uniform magnetic field to arise from gauge potentials having the form
$A_\mu = \Big( - B (x_2 - X) , 0 , 0, 0 \Big)$, 
for example, 
where 
$X$
is a constant. 
In finite volume, 
results at 
$X = 0$
are no longer gauge equivalent to those at
$X \neq 0$. 
This inequivalence is best expressed in terms of gauge holonomies. 
In this example, 
the Wilson loop 
$W_1(x_2)$
now must appear as
$W_1(x_2) = \exp [- i Q B L ( x_2 - X) ]$. 
As a majority of our results will be expressed in terms of the Wilson loops, 
it is largely trivial to account for alternate ways to implement the external field. 
} 
there are two non-vanishing holonomies which are coordinate dependent,
namely
\begin{eqnarray}
W_1(x_2) 
&=&
e^{ - i Q B L x_2}
,\notag \\ 
W_2(x_1) 
&=& 
e^{i Q B L x_1}
.\end{eqnarray}  
Addressing effects of these holonomies on neutron correlation functions is one of the central goals of this work.

Let the hadron possessing charge 
$Q$
be described by a field 
$\phi (x)$, 
and further that the gauge covariant derivative acting on the field has the form
$D_\mu \phi = \partial_\mu \phi + i Q A_\mu \phi$.  
Accordingly when the gauge field transforms as 
$A_\mu \to A_\mu + \partial_\mu \alpha$
under a gauge transformation, 
the matter field has the corresponding transformation
$\phi \to e^{-i Q \alpha} \phi$. 
In this way, 
the gauge covariant derivative transforms in precisely the same way as the matter field,
$D_\mu \phi \to e^{- i Q \alpha} D_\mu \phi$. 
Given these transformation properties, 
the non-periodicity of the gauge potential exhibited in 
Eq.~\eqref{eq:nonP}
can be gauged away at the boundary. 
As a result, 
a periodic matter field will satisfy a modified boundary condition after such a gauge transformation is made. 
Carrying out the boundary gauge transformation leads to the boundary condition
\begin{equation}
\phi(x + L \hat{x}_j)
= 
e^{ - i Q \Lambda_j (x)}
\phi(x),
\label{eq:MPBC}
\end{equation}
in accordance with gauge invariance of the Euclidean action density. 
We refer to Eq.~\eqref{eq:MPBC} as a magnetic periodic boundary condition
(MPBC). 
For the gauge potential at hand, 
the matter field satisfies the MPBC:
$\phi( x + L \hat{x}_2 ) = W_2(x_1) \phi (x)$, 
with periodicity obeyed in the other two spatial directions. 
The interpretation of this MPBC is simple, 
each time the charged particle winds around the 
$x_2$-direction, 
it picks up a Wilson line in accordance with gauge covariance. 
Although the field 
$\phi(x)$
is periodic in the 
$x_1$-direction, 
the other non-vanishing holonomy, 
$W_1(x_2)$,
remains relevant as we show through explicit computation. 
It enters through the violation of translational invariance.

Consistency of the MPBC requires quantization of the magnetic field. 
Consider winding around once in each direction transverse to the magnetic field to arrive at the field
$\phi (x + L \hat{x}_1 + L  \hat{x}_2)$. 
As the field is periodic in the 
$\hat{x}_1$
direction, 
we have 
$\phi (x + L \hat{x}_1 + L  \hat{x}_2) = \phi(x + L \hat{x}_2) = W_2(x_1) \phi(x)$.
On the other hand, 
the field satisfies a MPBC in the 
$\hat{x}_2$
direction, 
which can be taken into account first,
$\phi (x + L \hat{x}_1 + L  \hat{x}_2) = W_2( x_1 + L) \phi(x + L \hat{x}_1) = W_2(x_1 + L) \phi(x)$.
Consequently 
the Wilson loops must satisfy a consistency condition
$W_2(x_1) W_2^\dagger (x_1 + L) = 1$, 
which translates into a restriction on the size of the magnetic field
\begin{equation}
Q B L^2 = 2 \pi N_\Phi
\label{eq:quant}
,\end{equation}
where the integer
$N_\Phi$
is the magnetic flux quantum of the torus. 
The quantization of the magnetic field naturally leads to a discrete 
$\mathbb{Z}_{N_\Phi}$
translational invariance.
For integers
$n \in [ 0, 1, \cdots, N_\Phi-1]$, 
we have invariance of the Wilson loops under the translations
\begin{eqnarray}
W_1 \left( x_2 + \frac{n}{N_\phi} L \right) &=& W_1 (x_2),
\notag \\
W_2 \left(x_1 + \frac{n}{N_\phi} L \right) &=& W_2 (x_1)
.\end{eqnarray}
Such discrete translational invariance is a constraint that must be satisfied by the finite volume neutron effective action.

\subsection{Charged Scalar Propagator}

In considering finite volume corrections to the neutron two-point function in chiral perturbation theory, 
we will require the finite volume propagator for the charged pion. 
On a spatial torus, 
the Euclidean action for a charged scalar field 
$\phi$ 
has the familiar Klein-Gordon form
\begin{equation}
S
= 
\int_0^L d\vec{x}
\int_{- \infty}^\infty dx_4
\,
\phi^\dagger(x) 
\left[
- D_\mu D_\mu 
+ m^2
\right]
\phi(x)
.\end{equation}
On account of Eq.~\eqref{eq:MPBC}, 
the time-ordered correlation function for the scalar field,
$G_{FV} (x',x) = \langle 0 | T \left\{ \phi(x') \phi^\dagger(x) \right\} | 0 \rangle$,
must  satisfy the following 
MPBCs
\begin{eqnarray}
G_{FV} ( x' + L \hat{x}_2, x ) 
&=& 
W_2(x'_1) 
G_{FV} ( x', x),
\notag \\
G_{FV} ( x', x + L \hat{x}_2) 
&=&
G_{FV} (x', x) 
W_2^\dagger (x_1)
\label{eq:GMPBC}
.\end{eqnarray}
The scalar two-point correlation function obeys periodic boundary conditions in the remaining two spatial directions.

In order to satisfy the MPBCs required on the charged scalar two-point function, 
we follow%
~\cite{Tiburzi:2012ks}
and
construct the finite volume propagator from magnetic periodic images of the infinite volume propagator. 
To satisfy magnetic periodicity, 
we must take
\begin{equation}
G_{FV} (x', x)
=
\sum_{\vec{\nu}}
[ W^\dagger_2 (x'_1)]^{\nu_2}
G_\infty ( x' + \vec{\nu} L, x)
\label{eq:GFV}
.\end{equation}
Here we employ a shorthand notation for the sum over all winding numbers, 
$\sum_{\vec{\nu}} 
\equiv 
\sum_{\nu_1 = - \infty}^\infty 
\sum_{\nu_2 = -\infty}^\infty
\sum_{\nu_3 = -\infty}^\infty$, 
and the function 
$G_\infty (x',x)$
is the infinite volume two-point function, 
which satisfies the Green's function equation
$\left[ - D'_\mu D'_\mu + m^2 \right] G_\infty (x', x) = \delta^{(4)} (x' - x)$. 
As a consequence, 
the finite volume two-point function satisfies an analogous equation
\begin{equation}
\left[ - D'_\mu D'_\mu + m^2 \right]
G_{FV} (x',x)
= 
\delta^{(3)}_L (\vec{x} \,' - \vec{x}\,) \delta( x'_4 - x_4)
\label{eq:FVGF}
,\end{equation}
where
$\delta_L(x - y)$
is the Dirac delta-function appropriate for variables having compact support, 
$x, y \in [0,  L]$.

Without knowing the explicit form of the infinite volume correlation function,  
the first MPBC in Eq.~\eqref{eq:GMPBC} can easily be demonstrated by reindexing the sum over magnetic periodic images. 
Notice that despite the coordinate dependence introduced by the Wilson loops
$W_2^\dagger(x'_1)$ 
in Eq.~\eqref{eq:GFV}, 
the finite volume two-point function remains periodic in the 
$x_1$-direction. 
Demonstrating that the second MPBC is satisfied by 
Eq.~\eqref{eq:GFV}
requires the explicit form of the infinite volume correlation function, 
which appears as%
~\cite{Tiburzi:2008ma}
\begin{eqnarray}
G_\infty (x', x)
&=& 
\frac{1}{2} 
\int_0^\infty ds
\int \frac{d \bm{k}}{(2 \pi)^3}
e^{ i \bm{k} \cdot ( \bm{x}' - \bm{x} )}
e^{ - \frac{1}{2} s \cM^2}
\notag \\
&& \phantom{sp} \times
\Big\langle 
x'_2 - \frac{k_1}{Q B}, s 
\Big |
x_2 - \frac{k_1}{QB}, 0
\Big\rangle
\label{eq:SHOprop}
.\end{eqnarray} 
In this expression, 
we use the notation
$\int d \bm{k}$
for the integration
$\int_{- \infty}^\infty dk_1 \int_{- \infty}^\infty dk_3 \int_{-\infty}^\infty dk_4$, 
and the parameter 
$\cM$
is defined by 
$\cM^2 = m^2 + k_3^2 + k_4^2$. 
The bracketed quantity is the quantum mechanical propagator for the simple harmonic oscillator
\begin{eqnarray}
&&\langle x', t' | x, t \rangle 
= 
\langle x' | e^{- \Delta t \, H } | x \rangle  
\notag \\
&& \phantom{space}
= 
\sqrt{\frac{Q B}{2 \pi \sinh (Q B \Delta t) }}
\exp
\Bigg[
- \frac{Q B}{2 \sinh ( Q B \Delta t)}
\notag \\
&& \phantom{spacing}
\times
\Big\{ 
( x'^2 + x^2 ) 
\cosh (Q B \Delta t)
-
2 x' x
\Big\}
\Bigg]
,\end{eqnarray}
where 
$\Delta t = t' - t$
is the Euclidean time difference with 
$\Delta t > 0$
understood, 
and 
$H = \frac{1}{2} p_x^2 + \frac{1}{2} ( Q B x)^2$
is the simple harmonic oscillator Hamiltonian. 
The quantum mechanical propagator is manifestly an even function of
$Q B$, 
and thus we avoid superfluous 
$| Q B |$
notation.
The scalar propagator is invariant under
$Q B \to - QB$
accompanied by a mirror reflection in the 
$\hat{x}_1$-direction.
The propagator is also invariant under a simultaneous reflection of the 
$\hat{x}_1$
and
$\hat{x}_2$
directions. 
With the explicit form of the infinite volume two-point function, 
we can verify that the second MPBC is satisfied provided the magnetic field is appropriately 
quantized, 
as in Eq.~\eqref{eq:quant}.

To simplify computations below, 
it is useful to perform the 
$\bm{k}$
integrals appearing in Eq.~\eqref{eq:SHOprop}, 
which are each Gau\ss ian. 
The integration produces a one-dimensional, 
proper-time integral representation for the coordinate-space propagator
\begin{eqnarray}
G_\infty
(x',x)
&=&
e^{i Q B \Delta x_1 \ol x_2}
\int_0^\infty
\frac{ds}{(4 \pi s)^2}
\frac{Q B s}{\sinh Q B s}
e^{ - m_\pi^2 s}
\notag \\
&&
\times
\exp
\left[
- 
\frac{ QB \Delta \vec{x} \, {}^2_\perp }{4 \tanh Q B s} 
- 
\frac{\Delta x_3^2 + \Delta x_4^2}{4 s}
\right]
,\notag \\
\label{eq:Bprop}
\end{eqnarray}
with 
$\Delta x_\mu = x'_\mu - x_\mu$, 
and
$\ol x_\mu = \frac{1}{2} ( x'_\mu + x_\mu)$. 
The overall coordinate-dependent phase violates translational invariance, 
and accordingly vanishes when the external field is turned off. 
This phase factor leads to the appearance of Wilson loops 
$W_1(x_2)$
in finite volume computations, 
as well as 
an Aharonov-Bohm effect.

\subsection{Static Fermion Propagator}

The propagator for a charged fermion in a magnetic field can similarly be derived. 
We shall use the static approximation throughout,  
for which a simple form for the fermion propagator emerges.
We also carefully check that there are no effects from magnetic periodic images in the static limit.

The infinite volume static fermion propagator has the simple form
\begin{equation}
D_\infty (x', x)
= 
\delta^{(3)} (\vec{x} \, ' - \vec{x} \, ) \theta(x'_4 - x_4) 
\cP_+
,\end{equation}
where 
$\cP_+ = \frac{1}{2} ( 1 + \gamma_4 )$
is the positive parity projection matrix. 
As the fermion is charged, 
let us construct its finite volume propagator by taking a sum over magnetic periodic images
\begin{equation}
D_{FV} (x', x)
= 
\sum_{\vec{\nu}}
[ W^\dagger_2 (x'_1) ]^{\nu_2}
D_\infty (x' + \vec{\nu} L, x)
\label{eq:pprop}
,\end{equation}
so that the propagator satisfies MPBCs. 
Given the form of the infinite volume propagator, 
however,
the finite volume correlation function can be simplified dramatically.

Before attempting to simplify the heavy fermion correlation function, 
it is efficacious to relate the Dirac delta-function with compact support to a sum over non-compact delta-functions. 
This can be achieved through the Poission summation formula
\begin{equation}
\frac{1}{L}
\sum_{n = - \infty}^\infty
\delta ( k - 2 \pi n / L ) 
= 
\frac{1}{2 \pi} 
\sum_{\nu = - \infty}^\infty
e^{ i k L \nu}
.\end{equation}
The Dirac delta-function with compact support can be written as a Fourier decomposition over quantized momentum modes
\begin{equation}
\delta_L ( x - y)
=
\frac{1}{L}
\sum_{n = - \infty}^\infty
e^{ 2 \pi i n (x - y) / L}
.\end{equation} 
This sum over modes can then be converted into a winding number expansion by utilizing Poisson's formula. 
As a result, 
the compact Dirac delta-function is given by the alternate expression
\begin{equation}
\delta_L (x - y)
=
\sum_{\nu = - \infty}^\infty
\delta ( x + \nu L  - y)
,\end{equation}
which is written in terms of periodic images of the non-compact delta-function

Having recalled these useful properties of Dirac delta-functions, 
we are now ready to handle the case of the static fermion propagator with magnetic periodic images. 
From Eq.~\eqref{eq:pprop}, 
we can separate out the image contributions by writing the heavy fermion propagator in the form
\begin{equation}
D_{FV}(x',x) 
=
\cD_{FV} (\vec{x} \, ', \vec{x} \, ) \,
\theta(x'_4 - x_4) 
\cP_+
\label{eq:heavy}
,\end{equation}
where the winding number dependence has been relegated to the function
\begin{equation}
\cD_{FV} (\vec{x} \, ', \vec{x} \, )
=
\sum_{\vec{\nu}}
[W_2^\dagger (x'_1)]^{\nu_2}
\delta^{(3)} ( \vec{x} \, ' + \vec{\nu} L - \vec{x} \, )
\label{eq:pimage}
.\end{equation}
If the Wilson loops were absent from Eq.~\eqref{eq:pimage}, 
we would obviously have the simple relation
$\cD_{FV} (\vec{x} \, ', \vec{x} \, ) = \delta^{(3)}_L ( \vec{x} \, ' - \vec{x} \, )$. 
The presence of the Wilson loops, 
however, 
does not alter this relation. 
Physically this is quite sensible as the static fermion should remain at a fixed spatial location. 
To demonstrate that the relation holds in the presence of Wilson loops, 
we note that because the function 
$\cD_{FV} (\vec{x} \, ', \vec{x} \, )$
is written as a sum over non-compact Dirac delta-functions, 
we may utilize their continuous Fourier decomposition to arrive at the expression
\begin{equation}
\cD_{FV} (\vec{x} \, ', \vec{x} \, ) 
=
\sum_{\vec{\nu}}
\int \frac{d \vec{k}}{( 2 \pi)^3}
e^{ - i Q B L x'_1 \nu_2}
e^{ i \vec{k} \cdot ( \vec{x} \, ' + \vec{\nu} L  - \vec{x} \, )}
.\end{equation}
The sum over winding numbers can be converted into a momentum mode sum by utilizing the Poisson summation formula. 
This produces the discrete Fourier mode decomposition
\begin{eqnarray} 
\cD_{FV} (\vec{x} \, ', \vec{x} \, ) 
&=&
e^{ i Q B x'_1 ( x'_2 - x_2) }
\frac{1}{L^3}
\sum_{\vec{n}}
e^{ 2 \pi i \vec{n} \cdot ( \vec{x} \, ' - \vec{x} \, )}
\notag \\
&=&
\delta_L^{(3)}
(\vec{x} \, ' -  \vec{x} \, ) 
.\end{eqnarray}
The second equality holds because the overall phase factor becomes unity when multiplied by the delta-function.
Crucial to this observation is that 
$x'_2$
and
$x_2$
have compact support in 
$\cD_{FV} (\vec{x} \, ', \vec{x} \, )$.
The diligent reader will realize that the derivation of this relation is also required above to arrive at Eq.~\eqref{eq:FVGF}. 
Nonetheless, 
the static charged fermion propagator maintains a simple form in an external magnetic field.

\section{One-Loop Computation} 
\label{s:One} %

Having spelled out the physics of non-interacting particles in uniform magnetic fields on a torus, 
we are now in a position to calculate the neutron two-point function in a magnetic field including 
leading-order effects of pion-nucleon interactions. 
We first give a brief review of heavy nucleon chiral perturbation theory with 
strong-field power counting in Sec.~\ref{s:HNCPT}. 
This review will explain the necessary ingredients of the finite volume computation. 
As the computation of the neutron two-point function will be carried out in coordinate space, 
we provide a simple example in zero magnetic field to exhibit features that may be unfamiliar.
This zero-field example is presented in Sec.~\ref{s:EX}.  
Finally in Sec.~\ref{s:CMF}, 
we derive the spin-dependent and spin-independent terms in the neutron effective potential 
by integrating out charged pions winding around the torus.

\subsection{Heavy Nucleon Chiral Perturbation Theory}
\label{s:HNCPT}

The dynamics of the neutron at low-energies can be described in terms of an effective field theory, 
which is chiral perturbation theory. 
The main ingredient of chiral perturbation theory is the symmetry breaking pattern of QCD. 
For two massless quark flavors, 
the QCD action maintains an 
$SU(2)_L \otimes SU(2)_R \otimes U(1)_B$
symmetry that is spontaneously broken to 
$SU(2)_V \otimes U(1)_B$
by the formation of the chiral condensate. 
The resulting Goldstone manifold can be parameterized by a field 
$\Sigma$
that lives in the coset 
$SU(2)_L \otimes SU(2)_R / SU(2)_V$. 
The Goldstone modes are non-linearly realized in 
$\Sigma$:
\begin{equation}
\Sigma  
=
\exp \left( 2 i \varphi / f \right), 
\end{equation}
with the pions contained in 
$\varphi$ 
as
\begin{equation}
\varphi 
= 
\begin{pmatrix}
\frac{1}{\sqrt{2}} \pi^0 & \pi^+
\\
\pi^- & - \frac{1}{\sqrt{2}} \pi^0
\end{pmatrix}
.\end{equation}
Our conventions are such that the pion decay constant 
$f$
has value
$f \approx 130 \, \texttt{MeV}$. 
Under a chiral transformation 
$(L, R) \in SU(2)_L \otimes SU(2)_R$, 
we have the coset transformation 
$\Sigma \to L \Sigma R^\dagger$. 
The effective theory of pions can be constructed by writing down the most general
chirally invariant Lagrangian.

In constructing the chiral Lagrange density, 
one must consider additional sources of symmetry breaking. 
In an external magnetic field, 
for example, 
chiral symmetry is explicitly broken to the product of diagonal subgroups,
$SU(2)_L \otimes SU(2)_R \to U(1)_L \otimes U(1)_R$. 
Non-vanishing quark masses explicitly break chiral symmetry down to the vector subgroup. 
With these two sources of explicit symmetry breaking taken into account, 
the resulting low-energy theory maintains only a 
$U(1)_V \otimes U(1)_B$
symmetry. 
Accounting for this pattern of spontaneous and explicit symmetry breaking, 
the chiral Lagrange density has the form
\begin{equation}
\cL
=
\frac{f^2}{8}
\Tr \left( D_\mu \Sigma^\dagger D_\mu \Sigma \right)
- 
\frac{\lambda}{2}
\Tr \left( m_Q [ \Sigma^\dagger + \Sigma ] \right)
.\end{equation}
Appearing above is the quark mass matrix
$m_Q = \diag ( m_u, m_d )$, 
and we will work in the limit of strong isospin symmetry, 
$m_u = m_d = m$. 
The action of the covariant derivative on the coset field is specified by
\begin{equation}
D_\mu \Sigma = \partial_\mu \Sigma + i A_\mu \left[ \cQ, \Sigma \right]
,\end{equation}
where the quark electric charges appear in the matrix
$\cQ = e \diag \left( \frac{2}{3}, - \frac{1}{3} \right)$. 
The parameter
$- \lambda$
is the chiral limit value of the chiral condensate.

%
\smallskip
\begin{figure}
\epsfig{file=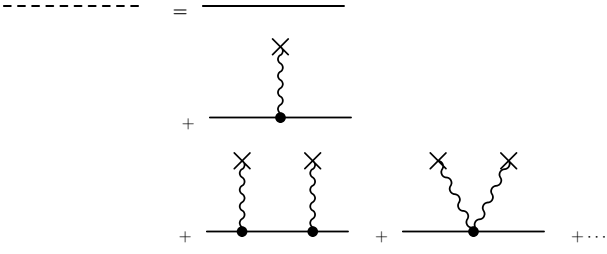,width=8cm}
\caption{\label{f:prop} 
        Charged pion propagator in an external magnetic field. 
        Depicted are the $\cO(p^2)$ couplings to the magnetic field which must 
        be summed to derive the propagator in strong-field power counting.}%
        \end{figure}
%

In writing the chiral Lagrange density, 
we have only included the lowest-order terms. 
We define these terms to scale as 
$\cO(p^2)$, 
where 
$p$
represents a small dimensionless number. 
The power counting we employ hence assumes that 
\begin{equation}
\frac{m_\pi^2}{\Lambda_\chi^2}
\sim
\frac{k^2}{\Lambda_\chi^2}
\sim
\frac{e F_{\mu \nu}}{\Lambda_\chi^2}
\sim
\frac{(e A_\mu)^2}{\Lambda_\chi^2} 
\sim
p^2
\label{eq:PC}
,\end{equation}
where 
$\Lambda_\chi \sim 2 \sqrt{2} \pi f$
is the chiral symmetry breaking scale. 
In this counting,  
$m_\pi$ 
is the pion mass, 
which satisfies the Gell-Mann--Oakes--Renner relation 
$f^2 m_\pi^2 = 4 \lambda m$.
The first condition of the power counting is implicitly a restriction on the size of the quark mass, 
which should be well satisfied for the up and down quarks.  
The second condition of the power counting concerns the typical momentum
$k$. 
In finite volume, 
the available momentum modes of a free particle are quantized in the form
$\vec{k} = \frac{2 \pi}{L} \vec{n}$, 
and thus the second condition implies the restriction to modes that satisfy
$\vec{n} \, {}^2 / 2 ( f L)^2 \sim p^2$. 
Combining the first and second conditions, 
we have the 
$p$-regime constraint%
~\cite{Gasser:1987zq}, 
namely
$m_\pi L \sim 2 \pi | \vec{n} | \gg 1$, 
for 
$| \vec{n} | \neq 0$. 
This constraint ensures that the pion Compton wavelength remains relatively large compared to the size of the torus. 
When this condition is not met, 
the pion zero modes become strongly coupled
and the chiral condensate will be substantially depleted%
~\cite{Gasser:1987ah}. 
The third condition of the power counting concerns the external magnetic field, 
whose strength must be weak compared to the chiral symmetry breaking scale.  
This condition treats the magnetic field as strong;
because, 
combining with the first condition, 
we have 
$|e B| / m_\pi^2 \sim 1$. 
Determination of corrections to the chiral condensate in this regime has been carried out in%
~\cite{Cohen:2007bt}.
Combining the second and the third conditions, 
we see the power counting supports many values of the magnetic flux quantum, 
$N_\Phi \sim 2 \pi | \vec{n} |^2 \gg 1$. 
The final condition of the power counting allows for the holonomies to appear at leading order. 
The argument of the Wilson loops have the general form 
$e B L x_\perp$, 
where 
$x_\perp = ( x_1, x_2)$
refers to either of the coordinates transverse to the direction of the magnetic field. 
By combining the second and final conditions, 
we arrive at 
$(e B x_2)^2 \sim k^2$, 
which translates into the condition
$| e B | L x_2 \sim 2 \pi | \vec{n} | \gg 1$. 
By the cubic symmetry of the torus, 
we must have 
$| e B | L x_\perp \gg 1$, 
too. 
In this counting, 
the Wilson loops are not amenable to perturbative expansion in the strength of the magnetic field.

In the strong-field, $p$-regime power counting, 
the neutral pion propagator retains its Klein-Gordon form. 
The charged pion propagator requires the summation of charge couplings as mandated by 
Eq.~\eqref{eq:PC}, 
see Fig.~\ref{f:prop} for a graphical depiction of the charged pion propagator. 
The leading-order charged pion Lagrange density appears as
\begin{equation}
\cL 
= D_\mu \pi^- D_\mu \pi^+ + m_\pi^2 \, \pi^- \pi^+ 
,\end{equation}
and thus the charged pion propagator has the form determined above in 
Eq.~\eqref{eq:GFV}.
This propagator,
$G_{FV}(x',x)$,
is the crucial new ingredient for the finite volume computation of the neutron two-point function.

The Lagrange density for a nucleon in an external field has the form 
\begin{equation}
\cL
=
\ol N 
\left(
\gamma_\mu 
D_\mu 
+ 
M
\right) N
,\end{equation}
where 
$N$ is the nucleon isodoublet, 
$N = \begin{pmatrix} p \\ n \end{pmatrix}$.  
We treat the mass
$M$
as the same order as the chiral symmetry breaking scale
$\Lambda_\chi$. 
In order to have a low-energy expansion, 
we must make a field redefinition to eliminate the mass term%
~\cite{Jenkins:1990jv}. 
To this end, 
we use the heavy nucleon field
$N_v$
defined by 
\begin{equation}
N_v(x)
= 
e^{M x_4} \cP_+ 
N(x)
.\end{equation}
With this redefinition, 
derivatives acting on the heavy nucleon field produce a residual momentum 
$k$ 
that can be treated as small, 
$k / M \sim \cO(p)$
in the power counting.
The relevant terms of the 
$\cO(p)$
nucleon chiral Lagrange density appear as
\begin{equation}
\cL 
=
\ol N_v \cD_4 N_v
+ 
g_A 
\ol N_v \vec{\sigma} \cdot \vec{\cA} \, N_v
,\end{equation}
where 
$\vec{\sigma}$
are the Pauli spin matrices, 
and 
$g_A \sim 1.25$
is the nucleon axial charge. 
In practice, 
the nearby 
delta-resonance makes sizable contributions to nucleon properties. 
Inclusion of these resonances is described in Appendix~\ref{s:D}. 
We omit the 
delta resonance in the main text only to keep the discussion as simple as possible. 
Electromagnetism is coupled into the pion-nucleon Lagrange density through the axial 
$\cA_\mu$
and
vector
$\mathcal{V}_\mu$
fields of pions. 
These are given by
\begin{eqnarray}
\mathcal{V}_\mu
&=& 
i Q A_\mu
+ 
\frac{1}{2 f^2}
\left(
\phi D_\mu \phi
- 
D_\mu \phi \, \phi 
\right)
+ 
\cdots, 
\notag
\\
\cA_\mu 
&=&
\frac{1}{f} D_\mu \phi + \cdots
,\end{eqnarray}
where terms of order 
$p^2$
and higher have been dropped. 
The action of the chirally covariant derivative 
$\cD_\mu$
on the nucleon field is specified by
\begin{equation}
(\cD_\mu)_i 
= 
\partial_\mu N_i
+
(\mathcal{V}_\mu)_i {}^{i'}
N_{i'}
+ 
\Tr \left( \mathcal{V}_\mu \right)
N_i
.\end{equation}

There are additional local interactions that are required to determine the neutron two-point function to 
$\cO(p^3)$. 
The magnetic field independent term at 
$\cO(p^2)$
is the nucleon mass operator, 
which gives rise to linear quark mass dependence of the nucleon mass away from the chiral limit.  
There are also magnetic field dependent operators, 
which are simply the isosinglet and isovector magnetic moment operators
\begin{equation}
\cL 
=
\frac{e}{2 M}
\left[
\kappa_0
\ol N_v \vec{\sigma} \cdot \vec{B}\,  N_v
+
\kappa_1
\ol N_v \vec{\sigma} \cdot \vec{B} \, \tau^3 N_v
\right]
\label{eq:magmom}
,\end{equation}
and also count as 
$\cO(p^2)$. 
Above we demonstrate that there is no effect from magnetic periodic images in the static limit. 
As a result, 
the propagators of both nucleons are given by 
$D_{FV}(x',x)$
in 
Eq.~\eqref{eq:heavy}.

\subsection{Example in Zero Magnetic Field}
\label{s:EX}

Before computing the neutron two-point function in an external magnetic field, 
it is efficacious to show how the finite volume computation of the nucleon mass proceeds in coordinate space. 
A salient feature of the coordinate space approach is that the propagators are written in a winding number expansion from the outset.

%
\begin{figure}
\epsfig{file=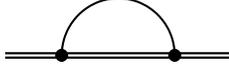,width=3cm}
\caption{\label{f:sunset} 
        Pion loop contribution to the nucleon mass in zero magnetic field. 
        The single line represents the free pion propagator, while double lines represent nucleon propagators. 
        }%
        \end{figure}
%

The one-loop sunset diagram for the nucleon two-point function is shown in Fig.~\ref{f:sunset}. 
Evaluating this diagram gives rise to the leading volume dependence of the nucleon mass. 
In coordinate space, 
the correction to the two-point function has the form
\begin{eqnarray}
\delta D_{FV} (x',x)
&=&
C 
\sumint
D_{FV}(x',y)
\sigma_i
D_{FV}(y,z)
\sigma_j
\notag \\
&& \phantom{sp}
\times
\left[
\frac{\partial^2}{\partial y_i \partial z_j}
G^{(0)}_{FV}(y,z)
\right]
D_{FV}(z,x)
,\notag \\
\label{eq:mess}
\end{eqnarray}
where 
$C$
is a dimensionful constant given by 
$\frac{3}{2} g_A^2 / f^2$. 
The factor of 
$\frac{3}{2} = \frac{1}{2} + 1$
takes into account both the neutral pion loop
($\frac{1}{2}$)
and the charged pion loop 
($1$).
The sum/integral notation reflects the 
$\mathbb{T}^3 \times \mathbb{R}$
spacetime
\begin{equation}
\sumint
=
\int_0^L d \vec{y} \int_{- \infty}^\infty dy_4
\int_0^L d \vec{z} \int_{- \infty}^\infty dz_4
.\end{equation}
For this zero-field example, 
the pion propagator is that of a free scalar
\begin{equation}
G_{FV}^{(0)}
(x',x)
=
\sum_{\vec{\nu}}
G_\infty^{(0)}
(x' + \vec{\nu} L,x)
,\end{equation}
with the infinite volume propagator
\begin{equation}
G^{(0)}_\infty (x',x) 
= 
\int_0^\infty \frac{ds}{ (4 \pi s)^2}
e^{ - m_\pi^2 s} 
e^{ - (x'-x)^2 / 4 s}
.\end{equation}
This familiar proper-time integral representation for the propagator emerges from the zero-field limit of 
Eq.~\eqref{eq:Bprop}.

To compute the nucleon mass shift from 
Eq.~\eqref{eq:mess}, 
we first amputate the external legs. 
In coordinate space, 
this is easy to achieve because propagators for the external legs are Green's functions satisfying the equation
$\frac{\partial}{\partial x'_4} D_{FV} (x',x) = \delta_{L}^{(3)} (\vec{x} \, ' - \vec{x} \, ) \delta(x'_4 - x_4)$. 
Defining the amputated contribution to the two-point function as
$[\delta D_{FV}(x',x)]_{\text{amp}} 
\equiv
- \frac{\partial^2}{\partial x'_4 \partial x_4}
\delta D_{FV} (x'x)$, 
we see that the amputated sunset diagram is
\begin{equation}
[\delta D_{FV}(x',x)]_{\text{amp}} 
=
C \,
D_{FV}(x',x)
\vec{\nabla}' \cdot \vec{\nabla}
G^{(0)}_{FV}(x',x)
,\end{equation} 
upon simplifying the spin structure.
Corrections to the nucleon mass are identified by projecting onto zero residual nucleon energy, 
$P_4 =0$.%
\footnote{
Away from this limit, 
one must deal with the first-order correction which is linear in 
$P_4$. 
This leads to the wavefunction renormalization factor, 
but is not needed to the order we are working in the chiral expansion
as it gives rise to terms at 
$\cO(p^4)$. 
} 
In general, 
we can write this projection in terms of an effective potential
\begin{equation} \label{eq:recipe}
\int_{- \infty}^\infty 
d(x'_4 - x_4) 
[\delta D_{FV}(x',x)]_{\text{amp}} 
=
-
V(\vec{x} \, ', \vec{x} \, )
.\end{equation}
The effective potential then appears as a correction to the nucleon Lagrangian
\begin{equation}
L(x_4) 
= 
\int_0^L d\vec{x} \,'
\int_0^L d\vec{x} \,\,
\ol N_v(\vec{x} \, ', x_4)
V(\vec{x} \, ', \vec{x} \, )
N_v( \vec{x} \, , x_4)
.\end{equation}
We have not appealed to spatial symmetries to simplify the form of the effective potential. 
Indeed for the simple example at hand, 
the potential is local, 
and otherwise coordinate independent. 
In an external magnetic field, 
however, 
we will find that the effective potential remains local, 
but is coordinate dependent.

Carrying out the projection in Eq.~\eqref{eq:recipe}, 
we indeed find a local potential, 
$V(\vec{x} \, ' , \vec{x} \, ) 
= 
\Delta M \,
\delta^{(3)}_L 
(\vec{x} \, ' - \vec{x} \, )$, 
with the constant 
$\Delta M$
given by
\begin{eqnarray}
\Delta M
&=&
C 
\frac{m_\pi^2 \sqrt{\pi}}{(4 \pi)^2}
\sum_{\vec{\nu}}
\int_0^\infty \frac{ds}{s^{3/2}}
e^{- m_\pi^2 s}
e^{ - \frac{\vec{\nu} \, {}^2 L^2}{ 4 s}}
,\end{eqnarray}
having utilized a proper-time integration by parts.
In this simple example, 
the constant
$\Delta M$ 
is just the correction to the nucleon mass. 
Separating out the infinite volume contribution,
$\Delta M_\infty$, 
which arises form the sector of zero winding number
$\vec{\nu} = \vec{0}$, 
we have 
\begin{equation}
\Delta M 
= 
\Delta M_\infty + \Delta M(L)
,\end{equation}
where the infinite volume contribution after dimensional regularization results in the well-known expression%
~\cite{Jenkins:1991ts,Bernard:1993nj}
\begin{equation}
\Delta M_\infty 
=
-
\frac{3 g_A^2}{16 \pi f^2}
m_\pi^3
\label{eq:infinite}
,\end{equation}
and the finite volume correction is
\begin{equation}
\Delta M(L)
=
\frac{3 g^2_A m^2_\pi}{16 \pi f^2}
\sum_{\vec{\nu} - \{ \vec{0} \} }
\frac{
e^{ 
- 
|\vec{\nu}| m_\pi L}}
{
|\vec{\nu}| 
L 
}
.\end{equation}
This expression for the finite volume correction to the nucleon mass is that derived in~\cite{Beane:2004tw}.
Having recovered familiar results from the coordinate-space approach, 
we are now ready to pursue the computation in an external magnetic field.

\subsection{Computation in Magnetic Field}
\label{s:CMF}

\begin{figure}
\epsfig{file=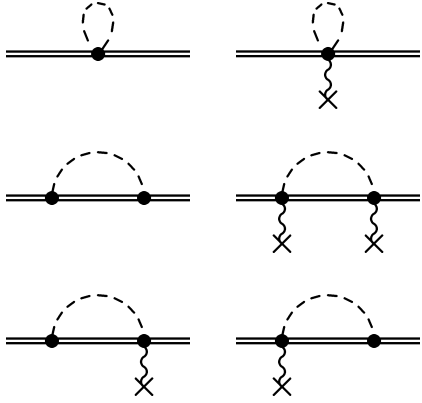,width=5.5cm}
\caption{\label{f:sunsetB} 
        Charged pion loop contributions to the neutron two-point function in an external magnetic field. 
        Dashed lines represent charged pion propagators in an external magnetic field, 
        while double lines represent nucleon propagators.
        The wiggly lines represent explicit couplings to the external field.  
        }%
        \end{figure}

To compute the 
$\cO(p^3)$
correction to the neutron effective action, 
we must evaluate the contributions depicted diagrammatically in 
Fig.~\ref{f:sunsetB}. 
The diagrams in the first row vanish, 
and the remaining four diagrams are related by gauge invariance. 
As a result, 
we can express the four sunset diagrams as one contribution by utilizing the gauge covariant derivative at each pion-nucleon vertex.  
Amputating the propagators of external legs produces 
\begin{equation}
[\delta D_{FV}(x',x)]_{\text{amp}} 
=
C \,
\sigma_i 
D_{FV}(x',x)
\sigma_j
D'_i D_j
G_{FV}(x',x)
\label{eq:amp}
,\end{equation} 
where the parameter 
$C$
now arises solely from the charged pion loop, 
and is accordingly given by
$C = g_A^2 / f^2$. 
This amputated correction must be projected onto vanishing neutron residual energy, 
$P_4 = 0$, 
to derive the neutron effective potential. 
There are both spin-independent and spin-dependent contributions to the effective potential.

%
%

\subsubsection{Spin-Independent Term}

The spin-independent part can be evaluated by considering how the action 
of 
$\vec{D}' \cdot  \vec{D}$
on 
$G_{FV} (x',x)$
differs from that of 
$- \vec{D}' {}^2$. 
To this end, 
we write
\begin{equation}
\vec{D}' \cdot  \vec{D} = - \vec{D}' {}^{2} + \vec{D}' \cdot  \vec{D}_+
,\end{equation}
where 
$\vec{D}_+ = \vec{D}' + \vec{D}$.  
From the expression for the propagator in position space,
Eqs.~\eqref{eq:GFV} and \eqref{eq:Bprop},
we see the only terms that can contribute to 
$\vec{D}_+$
are the phase factors. 
This is not surprising because it is precisely these factors that break translational invariance down to the discrete 
$\mathbb{Z}_{N_\Phi}$
subgroup. 
As translational invariance is preserved in the 
$\hat{x}_3$-direction, 
we have
$(D_+)_3 = 0$.

Now consider the derivatives acting in the 
$\hat{x}_2$-direction. 
We have simply
\begin{equation}
(D_+)_2 \,
G_\infty (x',x) 
= 
i Q B \Delta x_1 \,
G_\infty (x',x) 
.\end{equation}
A slightly more involved computation is required for the derivatives acting in the
$\hat{x}_1$-direction. 
On the 
$\vec{\nu} \, {}^\text{th}$
image,
we find that
\begin{eqnarray}
&&(D_+)_1 \, 
[W^\dagger_2(x'_1)]^{\nu_2}
G_\infty (x' + \vec{\nu} L,x)
\notag \\
&&\phantom{spa}
=
- i  Q B ( \Delta x_2 + \nu_2 L )
\, 
[W^\dagger_2(x_1)]^{\nu_2}
G_\infty (x' + \vec{\nu} L,x)
.
\notag \\
\end{eqnarray} 
Consequently on the 
$\vec{\nu} \, {}^\text{th}$
image, 
we are led, 
after some cancellations within the proper-time integral, 
to the equivalence
\begin{equation}
\vec{D}' \cdot \vec{D}_+ 
\longrightarrow
- \frac{1}{2} ( Q B L \vec{\nu}_\perp)^2
.\end{equation}
To evaluate the action of 
$\vec{D}' {}^2$, 
we write it in two terms, 
$\vec{D}' {}^2 = D'_\mu D'_\mu - (D'_4)^2$. 
For the former term, 
we can utilize the Green's function equation;
while for the latter term, 
it vanishes after projecting the amputated diagram onto definite 
$P_4 = 0$.

Assembling results of these observations, 
we can deduce the spin-independent neutron effective potential. 
To exhibit various features, 
we write the result in the form 
\begin{equation}
V_0 ( \vec{x} \, ', \vec{x} \, )
= 
\delta_L^{(3)} ( \vec{x} \, ' - \vec{x} \, )
\left[
E_0
+
V_0
(\vec{x}_\perp)
\right]
\label{eq:VeeZero}
.\end{equation}
Locality emerges due to the static limit of the proton propagator. 
The coordinate independent piece
$E_0$
appearing above is the spin-independent neutron energy shift in an external magnetic field. 
This term arises in infinite volume. 
Subtracting off the charged pion contribution to the nucleon mass appearing in 
Eq.~\eqref{eq:infinite}, 
we have
\begin{eqnarray}
E_0
=
\frac{g_A^2 m_\pi^2 \sqrt{\pi}}{(4 \pi f)^2}
\int_0^\infty \frac{ds}{s^{3/2}}
e^{ - m_\pi^2 s}
\left( 
\frac{e B s}{\sinh e B s}
-
1
\right)
\label{eq:strong0}
.\end{eqnarray}
In weak magnetic fields, 
$|e B| / m_\pi^2 \ll 1$, 
we can expand the neutron energy shift
$E_0$
in the form 
\begin{equation}
E_0 = - \frac{1}{2} 4 \pi \beta_M B^2 + \cdots,
\end{equation}
where 
$\beta_M$
is the magnetic polarizability given by 
$\beta_M = \frac{e^2}{4 \pi} \frac{g_A^2}{96 \pi f^2 m_\pi}$, 
and agrees with the known value from chiral perturbation theory%
~\cite{Bernard:1991rq}. 
The strong field result in Eq.~\eqref{eq:strong0} treats the magnetic field non-perturbatively, 
$|e B |/ m_\pi^2 \sim 1$, 
and was originally derived in~\cite{Tiburzi:2008ma}.%
\footnote{
Unfortunately there are mismatched factors of 
$2$
in 
Eqs.~(95) and (96) of~\cite{Tiburzi:2008ma}. 
These transcription errors arose from incomplete rescaling of the proper-time by a factor of 
$2$ 
in Eqs.~(A.5) and (A.7).
None of the other equations suffer from this error. 
} 
As a matter of curiosity, 
the Laplace transform required for
$E_0$
can be expressed in terms of well-known functions, 
namely
\begin{equation}
E_0 
= 
\frac{g_A^2 m_\pi^2 }{(4 \pi f)^2}
\sqrt{\pi | e B|}
\,
\mathcal{J} \left( \frac{m_\pi^2}{| e B|} \right)
\label{eq:E0nonP}
,\end{equation}
with
\begin{eqnarray}
\mathcal{J} (x)
&=&
\sqrt{2 \pi}
\left[
\sqrt{2 x} 
+ 
\zeta 
\left( 
\frac{1}{2}, 
\frac{x+1}{2}
\right)
\right]
\label{eq:Jzeta}
,\end{eqnarray}
where
$\zeta$
is the generalized zeta-function.


%
\begin{figure}
\begin{flushleft}
\epsfig{file=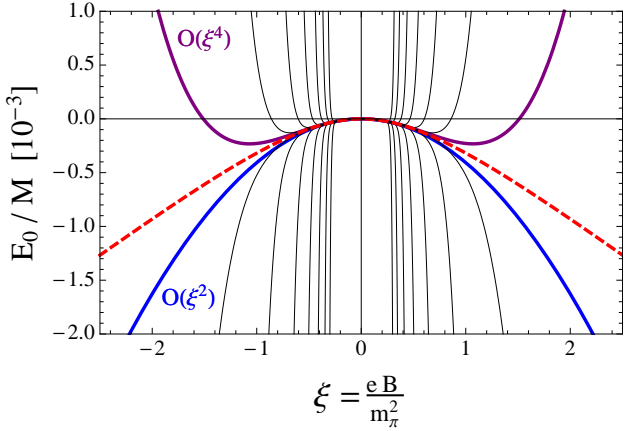,width=8.25cm}
\caption{\label{f:E0} 
Spin-independent energy shift of the neutron as a function of magnetic field. 
The first fifteen non-vanishing perturbative approximations to the neutron energy in powers of the magnetic field are compared with the non-perturbative result given in Eq.~\eqref{eq:E0nonP}. 
The dashed curve shows the non-perturbative magnetic field result, 
while the quadratic and quartic approximations to the energy are labeled. 
                }%
                \end{flushleft}
        \end{figure}
%


In Fig.~\ref{f:E0}, 
the spin-independent neutron energy in 
Eq.~\eqref{eq:E0nonP}
is compared with its weak-field limit. 
In particular, 
we focus outside the perturbative regime, 
where the expansion parameter,
$\xi = e B / m_\pi^2$,
satisfies
$|\xi| \gtrsim 1$.
Magnetic fields of this size and larger are readily encountered on current-day lattices. 
Outside the perturbative regime, 
adding higher-order corrections worsens the agreement with the non-perturbative result. 
This is the expected behavior of asymptotic expansions. 
Notice, 
however, 
that the non-perturbative result can be reasonably well approximated as a quadratic plus a quartic correction. 
Hence a fit to lattice data that line up with the non-perturbative curve
will not necessarily be able to distinguish between a perturbative model, 
and more complicated zeta-function behavior. 
Because the leading-order term of an asymptotic expansion results in the best approximation outside the perturbative regime,
it may well be that the value of the quadratic coefficient determined from a perturbative model fit has a value close to the true leading-order coefficient. 
It should be noted that this procedure has a fundamental limitation that can only be addressed by comparing 
perturbative model fits to those using the expected non-perturbative behavior.

\begin{figure}
\epsfig{file=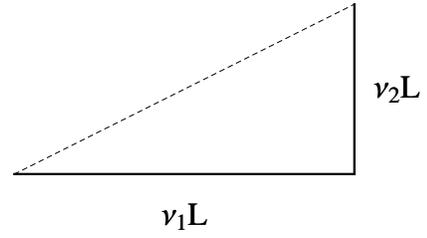,width=5.5cm}
\caption{\label{f:AB} 
        Depiction of charged pion winding in the plane transverse to the magnetic field. 
        The pion propagates a distance
        $\nu_1 L$
        in the 
        $\hat{x}_1$-direction, 
        and 
        $\nu_2 L$
        in the 
        $\hat{x}_2$-direction. 
        This propagation is accompanied by two Wilson loops. 
        By periodicity, 
        the pion ends up at a point equivalent to where it started, 
        and additionally must acquire an Aharonov-Bohm phase, 
        $e^{ i Q A B}$. 
        }%
        \end{figure}

The remaining term 
$V_0(\vec{x}_\perp)$
appearing in the spin-independent neutron energy,
Eq.~\eqref{eq:VeeZero}, 
is the effective potential for the neutron, 
and it arises in this computation from finite volume effects. 
This spin-independent potential has the form
\begin{eqnarray}
V_0(\vec{x}_\perp)
&=& 
\frac{g_A^2 \sqrt{\pi}}{(4 \pi f)^2}
\sum_{\vec{\nu} - \{\vec{0} \}}
(-)^{N_\Phi \nu_1 \nu_2}
[W^\dagger_1(x_2)]^{\nu_1}
[W^\dagger_2(x_1)]^{\nu_2}
\notag\\
&&
\times
\int_0^\infty \frac{ds}{s^{3/2}}
\frac{e B s}{\sinh e B s}
\left[ m_\pi^2 + \frac{1}{2} ( e B  \vec{\nu}_\perp L)^2 \right]
\notag \\
&&
\times
\exp
\left[ 
- m_\pi^2 s
- 
\frac{e B \vec{\nu}_\perp^{\, 2} L^2}{4 \tanh e B s} 
- 
\frac{\nu_3^2 L^2}{4 s}
\right]
.
\label{eq:V0}
\end{eqnarray}
Because this result depends on the Wilson loops, 
the effective potential obeys the discrete 
$\mathbb{Z}_{N_\Phi}$
magnetic translational invariance of the torus. 
It is amusing to note that the oscillating sign factor,
$(-)^{N_\Phi \nu_1 \nu_2}$,
naturally arises from the Aharonov-Bohm effect, 
$(-)^{N_\Phi \nu_1 \nu_2} = e^{i e B (\nu_1 L) (\nu_2 L) / 2}$, 
where, 
by periodicity,  
the flux penetrates a triangular region in the plane transverse to the magnetic field, 
see Fig.~\ref{f:AB}.

%
%

In the limit of vanishing magnetic field, 
the coordinate dependence of the effective potential disappears, 
and we recover the charged pion contribution to the finite volume neutron mass, 
$V_0(\vec{x}_\perp) \overset{B\to0}{=} \Delta M(L)$. 
In a non-vanishing magnetic field, 
the winding number sums can be cast in terms of Jacobi elliptic-theta functions. 
As the resulting expression for
$V_0(\vec{x}_\perp)$
is rather lengthy, 
we display the result in Appendix~\ref{s:A}. 
Various features of the result will be exhibited below in 
Sec.~\ref{s:Disc}.

\subsubsection{Spin-Dependent Term}

We now evaluate the spin-dependent term entering the one-loop expression for the neutron two-point function. 
Using the familiar properties of Pauli matrices, 
we see that the spin-dependent term in 
Eq.~\eqref{eq:amp}
is proportional to 
$i \vec{\sigma} \cdot (\vec{D}' \times \vec{D} )$. 
Hence whenever the action of 
$\vec{D}$
is the same as 
$- \vec{D}'$, 
there is a spin-dependent contribution of the form 
$- i \vec{\sigma} \cdot (\vec{D}' \times \vec{D}' )
= 
Q \, \vec{\sigma} \cdot \vec{B} 
= 
Q B \sigma_3$.
To this end, 
we write
\begin{equation}
i \vec{\sigma} \cdot (\vec{D}' \times \vec{D} )
=
Q B \sigma_3 + i \vec{\sigma} \cdot ( \vec{D}' \times \vec{D}_+)
,\end{equation} 
where the action of 
$\vec{D}_+$
on the 
$\vec{\nu} \, {}^\text{th}$
magnetic periodic image has been determined above. 
To simplify the cross product, 
we notice that 
$D'_3$
is proportional to 
$\Delta x_3$; 
and, 
in the 
$\vec{\nu} \, {}^\text{th}$
image, 
this displacement becomes 
$\nu_3 L$
which is odd under reindexing
$\nu_3 \to - \nu_3$. 
All of the remaining factors in propagators of magnetic periodic images are even in 
$\nu_3$. 
We can therefore eliminate 
$D'_3$
from the cross product. 
Because
$\vec{D}_+$
has only components transverse to the magnetic field, 
and the
$D'_3$
contribution is zero, 
the finite volume effect only allows the neutron spin to point in the direction of the magnetic field, 
as in infinite volume. 
Thus we have
\begin{equation}
i \vec{\sigma} \cdot ( \vec{D}' \times \vec{D}_+ )
\longrightarrow
i \sigma_3 [ D'_1 \,  (D_+)_2 - D'_2 \,  (D_+)_1]
.\end{equation}
In determining this contribution, 
we must not forget terms of the form 
$i \sigma_3 \Big( [ D'_1,  (D_+)_2]  - [D'_2,  (D_+)_1] \Big)$, 
which evaluate to 
$ - 2 Q B \sigma_3$.

Assembling the results of these observations produces the expression for the spin-dependent neutron effective potential. 
A convenient way to write the result is 
\begin{equation}
V_1 (\vec{x} \, ', \vec{x} )
=
e B \sigma_3 \, 
\delta_L^{(3)}
(\vec{x} \, ' - \vec{x} )
\left[
E_1
+ 
V_1 ( \vec{x}_\perp )
\right]
\label{eq:VeeOne}
,\end{equation}
where locality again emerges due to the static limit of the intermediate-state proton propagator. 
The coordinate independent piece,
$E_1$
determines the spin-dependent neutron energy in the magnetic field, 
which leads to a Zeeman energy splitting: 
$2 e B E_1$.
This contribution arises from the infinite volume limit, 
and can be written in the form
\begin{equation}
E_1 
= 
\frac{\kappa_n}{ 2 M}
+ 
\Delta E_1
\label{eq:strong1}
.\end{equation}
Here
$\kappa_n $
is the neutron magnetic moment, 
which includes the tree-level contribution, 
$\kappa_0 - \kappa_1$, 
from the magnetic moment operators in 
Eq.~\eqref{eq:magmom}, 
as well as the charged pion loop contribution.
The latter emerges from our computation in the sector of zero winding number, 
$\vec{\nu} = \vec{0}$, 
and is the contribution at linear order in the magnetic field. 
Using dimensional regularization, 
we see
\begin{equation}
\kappa_n 
= 
\kappa_0 - \kappa_1 
+ 
\frac{g_A^2 M m_\pi}{4 \pi f^2}
,\end{equation}
which is the standard result from chiral perturbation theory%
~\cite{Jenkins:1992pi,Meissner:1997hn}.
The contribution to the energy proportional to  
$\Delta E_1$
arises from treating the magnetic field as strong. 
It is given by the expression
\begin{equation}
\Delta E_1
=
- 
\frac{g_A^2}{( 4 \pi f)^2}
\sqrt{ \pi | e B |}
\mathcal{J} \left( \frac{m_\pi^2}{|e B|} \right)
\label{eq:E1nonP}
,\end{equation}
where 
$\mathcal{J}(x)$
is defined in Eq.~\eqref{eq:Jzeta}.


%
\begin{figure}
\begin{flushleft}
\epsfig{file=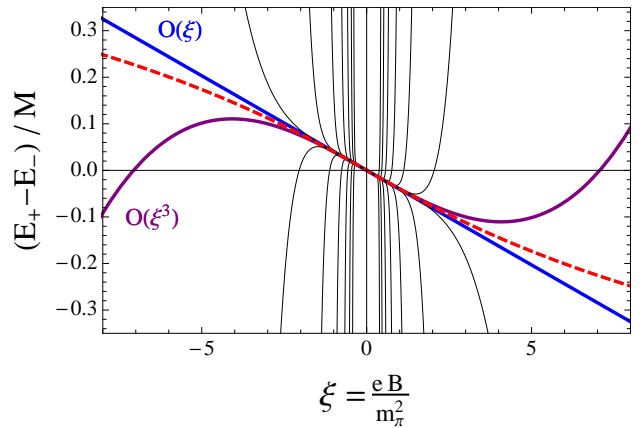,width=8.25cm}
\caption{\label{f:E1} 
Zeeman splitting of the neutron as a function of magnetic field. 
The first fifteen non-vanishing perturbative approximations to the Zeeman splitting in powers of the magnetic field are compared with the non-perturbative result which follows from Eq.~\eqref{eq:strong1}. 
The dashed curve shows the non-perturbative magnetic field result, 
while the linear and cubic approximations to the splitting are labeled. 
                }%
                \end{flushleft}
        \end{figure}
%


In Fig.~\ref{f:E1}, 
the neutron Zeeman splitting is plotted as a function of the magnetic field. 
The splitting is defined to be the energy difference between spin-polarized states normalized by the neutron mass. 
We use the physical neutron magnetic moment to generate this plot. 
As with the spin-independent case, 
our primary focus in the spin-dependent case is outside the perturbative regime, 
and  our consideration  is motivated by the size of uniform magnetic fields available on typical lattices. 
In the case of the Zeeman splitting, 
we consider values of the expansion parameter satisfying 
$|\xi| \sim 5$. 
For such magnetic fields, 
the perturbative expansion has broken down, 
as is evidenced by the figure. 
Despite this breakdown, 
the non-perturbative result follows a strikingly linear curve, 
with only slight curvature appearing as 
$|\xi|$
approaches 
$10$. 
Such field strengths, 
however, 
satisfy the relation
$| e B| /M m_\pi \sim 1$,
and our results are not reliable for these considerably large values of the magnetic field. 
It would be interesting to study whether this fortuitous behavior persists when recoil-order corrections are added. 
Nonetheless, 
fits to lattice data that follow the non-perturbative curve 
will not necessarily be able to distinguish between a perturbative model, 
and more complicated non-perturbative behavior. 
The figure suggests that the value of the magnetic moment deduced from fitting the non-perturbative curve to a straight line will not be far off from the actual value of the magnetic moment. 
While the leading-order term of an asymptotic expansion results in the best approximation outside the perturbative regime,
this approximation is uncontrolled with no guarantee that the leading-order term is even close to the non-perturbative result. 
Chiral perturbation theory suggests that the magnetic moment might be a fortuitous case, 
but care is needed to evaluate the behavior of the expansion in powers of the magnetic field.

The remaining term 
$V_1(\vec{x}_\perp)$
appearing in the spin-dependent neutron energy, 
Eq.~\eqref{eq:VeeOne},  
arises solely from finite volume effects. 
This spin-dependent neutron potential takes the form
\begin{eqnarray}
V_1(\vec{x}_\perp)
&=& 
\frac{g_A^2 \sqrt{\pi}}{(4 \pi f)^2}
\sum_{\vec{\nu} - \{\vec{0} \}}
(-)^{N_\Phi \nu_1 \nu_2}
[W^\dagger_1(x_2)]^{\nu_1}
[W^\dagger_2(x_1)]^{\nu_2}
\notag \\
&&
\times
\int_0^\infty \frac{ds}{s^{3/2}}
\frac{e B s}{\sinh e B s}
\left[
-
1
+ 
\frac{ e B \vec{\nu}^{\, 2}_\perp L^2}{2 \tanh e B s}  
\right]
\notag \\
&&
\times
\exp
\left[ 
- m_\pi^2 s
- 
\frac{e B \vec{\nu}_\perp^{\, 2} L^2}{4 \tanh e B s} 
- 
\frac{\nu_3^2 L^2}{4 s}
\right]
.\label{eq:V1}
\end{eqnarray}
Despite the explicit coordinate dependence, 
the potential maintains the discrete 
$\mathbb{Z}_{N_\Phi}$
magnetic translational invariance. 
As with the spin-independent potential, 
the oscillating sign factor present in this expression, 
$(-)^{N_\Phi \nu_1 \nu_2}$, 
reflects the Aharonov-Bohm effect. 
The winding
number sums can be recast in terms of Jacobi elliptic-theta functions, 
and we display this lengthy result in Appendix~\ref{s:A}.
Because the potential takes a rather complicated form, 
we discuss various features by first considering simplifying limits.

\section{Discussion and Conclusion} 
\label{s:Disc} %

\subsection{Discussion of Results}

Above we determine the effective potential for the neutron in a magnetic field by integrating out charged pions winding around the torus. 
The charged pion winding is accompanied by Wilson loops that reflect the non-trivial holonomy of the gauge field in accordance with gauge invariance. 
As the general result is rather complicated, 
we exhibit features by using various limits before explaining how to determine the finite volume effect more generally.

\subsubsection{Extreme Weak-Field Limit}

As the simplest limit to consider, 
let us imagine the limit of an extremely weak field,
$| e B| / m_\pi^2 \ll 1$
and
$| e B | L x_\perp \ll 1$. 
The latter constraint requires localization of the physics to the bulk of the lattice. 
In this limit, 
we could expand the Wilson loops perturbatively in the strength of the magnetic field;
although, 
we argue against this below.

As the spin-dependent potential starts out at linear order in the magnetic field, 
we can imagine taking the strict zero-field limit of the function 
$V_1(x_\perp)$,
which results in a coordinate-independent finite volume effect
\begin{eqnarray}
V_1(x_\perp)\Big|_{B = 0}
&=&
\frac{g_A^2 \sqrt{\pi}}{(4 \pi f)^2}
\sum_{\vec{\nu} - \{\vec{0} \}}
\int_0^\infty \frac{ds}{s^{3/2}}
\left[
-
1
+ 
\frac{\vec{\nu}^{\, 2} L^2}{3 s}  
\right]
\notag \\
&&
\times
\exp
\left[ 
- m_\pi^2 s
- 
\frac{\vec{\nu}^{\, 2} L^2}{4 s}
\right]
.\end{eqnarray}
In an extremely weak magnetic field, 
this contribution could be identified as the finite volume effect on the neutron magnetic moment. 
Performing the proper-time integration results in
\begin{equation}
\Delta \kappa_n(L)
=
\frac{g_A^2 M m_\pi}{6 \pi f^2}
\sum_{\vec{\nu} - \{\vec{0} \}}
\left[
1
-
\frac{1}{2 |\vec{\nu} \, | m_\pi  L}
\right]
e^ {- |\vec{\nu}\, | m_\pi L}
\label{eq:FVmoment}
.\end{equation}
This result for the finite volume modification to the magnetic moment differs from the one obtained in~\cite{Beane:2004tw}.%
\footnote{
For reference, 
the result claimed in~\cite{Beane:2004tw} for the neutron magnetic moment is given by
$\frac{g_A^2 M m_\pi}{12 \pi f^2}
\sum_{\vec{\nu} - \{\vec{0} \}}
\left[
1
-
\frac{2}{|\vec{\nu} \, | m_\pi  L}
\right]
e^ {- |\vec{\nu} \, | m_\pi L}$.
} 
That result, 
however, 
was obtained by taking the derivative
$\partial/\partial m_\pi^2$
of the finite volume correction to the nucleon mass. 
While infinite volume results are related by taking a quark mass derivative, 
the finite volume corrections are not. 
In a three-point function computation of the current matrix element, 
additional terms appear in finite volume due to the quantization of momentum transfer between initial- and final-state nucleons. 
These terms reflect the breaking of rotational invariance. 
A detailed critique of the computation of%
~\cite{Beane:2004tw} 
appears in%
~\cite{Tiburzi:2007ep}.
On the other hand, 
isospin twisted boundary conditions provide a method to overcome the limitation to quantized momentum transfer.
Finite volume corrections to the nucleon isovector magnetic moment with isospin twisted boundary conditions have been determined in%
~\cite{Tiburzi:2006px}.  
When one takes the limit of vanishing twist parameter
$\theta$, 
the finite volume correction to the neutron magnetic moment is precisely that appearing above in 
Eq.~\eqref{eq:FVmoment}.
This equivalence is demonstrated in Appendix~\ref{s:B}.

The situation is different with respect to the extreme weak-field limit of the spin-independent neutron potential. 
In the unpolarized case, 
the strict zero-field limit produces the finite volume correction to the neutron mass
(from the charged pion loop). 
We must then expand to second order in the magnetic field to arrive at new results. 
Potentially contained in such results would be the finite volume correction to the magnetic polarizability. 
At this order, 
however,
terms arising from expanding the holonomies will be present and these lead to coordinate dependence. 
Analogous results have been considered for pion two-point functions in perturbatively small external fields, 
see~\cite{Tiburzi:2008pa}.
The oscillating signs arising from Aharonov-Bohm phases, 
moreover, 
cannot be sensibly expanded even in an extremely weak magnetic field. 
We will avoid the temptation to expand any phase factors in powers of the magnetic field. 
These topological effects are inherently non-perturbative in the field strength, 
and properly reflect the physics on a torus.

\subsubsection{Magnetic Fields on a Large Lattice}

Simple expressions can be obtained by considering uniform magnetic fields on a finite lattice. 
We take the lattice spacing,
$a$,
to have the same value in directions transverse to the magnetic field,
with the remaining directions irrelevant to our discussion. 
As a result, 
the transverse coordinates can be indexed by lattice sites, 
$\vec{x}_\perp = a \, \vec{n}_\perp$.
The restriction to finite lattice spacing
leads to another possible quantization condition for the magnetic field
\begin{equation}
Q B = \frac{2 \pi N_\Phi}{a L}
\label{eq:otherB}
.\end{equation}
On typical lattices, 
such magnetic fields are prohibitively large; 
however, 
in the present context, 
they eliminate complications from the gauge holonomies.  
For magnetic fields obeying the quantization condition in 
Eq.~\eqref{eq:otherB}, 
we have
\begin{eqnarray}
W_1(x_2)
&=&
e^{- 2 \pi i N_\Phi x_2 / a}
=
1
,\notag \\
W_2(x_1)
&=&
e^{ 2 \pi i N_\Phi x_1 / a}
=
1
,\end{eqnarray}
because 
$\vec{x}_\perp / a$
indexes the lattice sites. 
The effect of topology is almost entirely eliminated. 
The Aharonov-Bohm phases still remain, 
but depend on the lattice volume
\begin{equation}
e^{i Q B L^2 \nu_1 \nu_2 / 2}
=
(-)^{N_\Phi \nu_1 \nu_2 L / a}
,\end{equation}
where 
$L / a$
is the number of lattice sites in a transverse direction. 
When the magnetic field is quantized according to 
Eq.~\eqref{eq:otherB}, 
the Aharonov-Bohm phases disappear on even-site lattices. 
In the following, 
we assume that the lattice has an even number of sites in the transverse directions for ease. 

Due to the absence of holonomies, 
the finite volume corrections in the spin-independent and spin-dependent cases are given simply by
$V_0 (\vec{0}_\perp)$
and
$V_1(\vec{0}_\perp)$, 
respectively. 
If we demand to identify finite volume corrections to the magnetic moment and magnetic polarizability from these results, 
then we must additionally assume the weak-field limit,
$|e B| / m_\pi^2 \ll 1$. 
Because the spin-dependent energy is already linear in the magnetic field, 
we can evaluate the finite volume correction
$V_1( \vec{0}_\perp)$
to zeroth order in the weak-field limit. 
This leads us to the finite-volume correction to the magnetic moment given above in Eq.~\eqref{eq:FVmoment}, 
provided the lattice has an even number of sites in the transverse directions.

We can additionally deduce the finite volume correction to the magnetic polarizability by expanding the spin-independent energy shift in powers of the magnetic field.
Because we temporarily adopt the quantization condition in Eq.~\eqref{eq:otherB}, 
there are no holonomies complicating the expansion in powers of the magnetic field.
Additionally the assumption of an even number of lattice sites removes Aharonov-Bohm phases.  
Writing the expansion of the spin-independent finite volume effect in the form 
\begin{equation}
V_0 (\vec{0}_\perp)
=
\Delta M (L)
- 
\frac{1}{2} 4 \pi \, \Delta \beta_M (L)
\, B^2
+ 
\cdots
\label{eq:weakV0}
,\end{equation}
we find the finite volume correction to the magnetic polarizability is given by
\begin{eqnarray}
\Delta \beta_M (L)
&=&
\frac{e^2}{4\pi}
\frac{g_A^2 m_\pi L^2}{144 \pi f^2}
\sum_{\vec{\nu} - \{ \vec{0} \} }
|\vec{\nu} |^2 
e^{ - | \vec{\nu} | m_\pi  L}
\notag \\
&& \times
\Bigg[ 
1
- 
\frac{21}{2}
\frac{1}{
| \vec{\nu} | m_\pi  L}
+ 
\frac{3}{2}
\frac{1}{(\vec{\nu} \, m_\pi L)^2}
\Bigg]
\label{eq:betaFV}
.\end{eqnarray}
We have been unable to relate this result to the one claimed in%
~\cite{Detmold:2006vu}. 
That analysis, 
however, 
neglected breaking of rotational invariance. 
It is conceivable that a computation of the Compton scattering tensor with twisted boundary conditions taken in the limit of vanishing twist angles will reproduce the finite volume effect in Eq.~\eqref{eq:betaFV}. 
An investigation along these lines is left to future work.

Blindly using the weak-field result to compute finite volume corrections to the magnetic polarizability yields unreasonably large corrections on the order of 
$\sim 500 \%$
at the physical pion mass, 
for a lattice size of
$L = 4 \, \texttt{fm}$. 
These volume effects can be softened to 
$\sim 50 \%$
by retaining terms of all orders in 
$e B / m_\pi^2$; 
however, 
the underlying problem concerns the size of magnetic fields arising from the quantization condition in
Eq.~\eqref{eq:otherB}
employed to derive the above finite-size effect. 
The strong-field condition, 
$| e B | / m_\pi^2 \sim 1$,
in conjunction with the field quantization in Eq.~\eqref{eq:otherB}, 
translates into the restriction
\begin{equation}
m_\pi L \sim \frac{2 \pi N_\Phi}{a  \, m_\pi} \sim 200
.\end{equation}
To arrive at this estimate for the required size of 
$m_\pi L$, 
we assume the smallest magnetic flux quantum possible for QCD,  
$N_\Phi = 3$, 
which is due to the fractional electric charges of quarks, 
and 
take a reasonable lattice spacing of 
$a = 0.1 \, \texttt{fm}$,
with a typical pion mass of 
$m_\pi = 200 \, \texttt{MeV}$. 
While the quantization condition in 
Eq.~\eqref{eq:otherB} 
has been utilized to arrive at desirably simple expressions for finite volume effects, 
namely those given in
Eqs.~\eqref{eq:FVmoment} and \eqref{eq:betaFV}, 
lattice practitioners should be wary of their use. 
Magnetic moments and magnetic polarizabilities do not have unique definitions on a torus, 
where rotational invariance is lost. 
As a consequence, 
finite volume corrections to these quantities must intrinsically depend on the lattice method employed in their calculation. 
Universal corrections are not possible except in extreme limits:
asymptotically large lattices where magnetic fields and momentum transfers are infinitesimal, 
or finite size lattices with infinitesimal twisted boundary conditions. 
These extreme limits provide an environment where conventional definitions of moments and polarizabilities can be realized, 
albeit not necessarily in practice.

\subsubsection{Beyond the Static Limit}

Having detoured to extreme cases, 
let us return to the strong-field power counting with the magnetic field quantized according to
Eq.~\eqref{eq:quant}. 
To evaluate the finite volume effects derived above, 
we must confront the Wilson loops that appear from charged pions winding around the torus. 
In turn, 
this requires us to go beyond the static limit in order to handle the coordinate dependence of the effective action. 
With the addition of the non-relativistic kinetic term, 
the neutron effective action takes the form
\begin{eqnarray}
\mathcal{L}_{\text{eff}}
&=&
N^\dagger
\Bigg[
\partial_4
- 
\frac{\vec{\nabla}_\perp^2}{2 M}
+ 
E_0
+  
V_0 ( \vec{x}_\perp)
\notag \\
&& \phantom{space}
+ 
e B \sigma_3
\left[
E_1 
+ 
V_1( \vec{x}_\perp)
\right]
\Bigg]
N
,\end{eqnarray}
where we have projected the neutron onto vanishing $\hat{x}_3$-component of momentum. 
With the coordinate dependence introduced by finite volume effects, 
however,
the transverse momentum is no longer a good quantum number. 
Beyond the static limit, 
the neutron propagator has an eigenfunction expansion in terms of eigenmodes of the above action. 
Consequently the Fourier projection of neutron correlation functions onto 
$\vec{P}_\perp = \vec{0}_\perp$
will receive contributions from the entire set of energy eigenstates. 
In lattice QCD, 
this complication occurs in addition to hadronic excited-state contributions, 
which is reminiscent of the Landau level problem for charged particles in magnetic fields, 
see%
~\cite{Tiburzi:2012ks}.
The effect of finite volume is one of the complicating features for neutron spectroscopy in lattice QCD with external fields.

To discuss effects of the coordinate-dependent potentials, 
we begin by isolating the coordinate dependence with a subtraction of images having 
$\vec{\nu}_\perp = \vec{0}_\perp$. 
To this end, 
we write the terms of the neutron effective action in the form
\begin{eqnarray}
V_0(\vec{x}_\perp)
&=&
\Delta M(L) + 
\Delta E_0(L) + 
U_0(\vec{x}_\perp), 
\notag 
\\ 
V_1(\vec{x}_\perp)
&=&
\Delta E_1(L) + U_1(\vec{x}_\perp),
\end{eqnarray}
where all of the coordinate dependence has been relegated to the residual potentials, 
$U_0(\vec{x}_\perp)$
and
$U_1(\vec{x}_\perp)$. 
These potentials arise from charged pions winding around the compact directions transverse to the magnetic field,
as such windings produce Wilson loops. 
In this way, 
the finite volume corrections  
$\Delta E_0(L)$
and
$\Delta E_1(L)$
arise solely from charged pions winding around the 
$\hat{x}_3$-direction. 
We additionally separate the magnetic field independent piece of the spin-independent finite volume effect. 
This is just the finite volume correction to the neutron mass, 
$\Delta M(L)$, 
arising from the charged pion loop. 
Because the coordinate-independent finite volume effects arise from winding only in the 
$\hat{x}_3$-direction, 
they are given by relatively simple expressions
\begin{eqnarray}
\Delta E_0(L)
&=&
\frac{g_A^2 m_\pi^2 \sqrt{\pi}}{(4 \pi f)^2}
\int_0^\infty \frac{ds}{s^{3/2}}
\left(
\frac{e B s}{\sinh e B s}
-1
\right)
\notag \\
&& \phantom{space}
\times 
e^{- m_\pi^2 s} 
[ \vartheta_3 (0, e^{- L^2 / 4 s}) -1]
\label{eq:E0}
,\end{eqnarray}
for the spin-independent case, 
and
\begin{eqnarray}
\Delta E_1(L)
&=&
-
\frac{g_A^2 \sqrt{\pi}}{(4 \pi f)^2}
\int_0^\infty \frac{ds}{s^{3/2}}
\frac{e B s}{\sinh e B s}
e^{ - m_\pi^2 s}
\notag \\
&& \phantom{spacingspa}
\times
[ \vartheta_3 (0, e^{- L^2 / 4 s}) -1]
\label{eq:E1}
,\end{eqnarray}
for the spin-dependent case. 
The Jacobi elliptic-theta functions appearing above are defined in 
Appendix~\ref{s:A}.

The residual potentials can both be written in the form
\begin{eqnarray}
U_j (\vec{x}_\perp)
&=&
\sum_{\vec{\nu} - \{ \vec{0}, \vec{0}_\perp \}}
[W^\dagger_1(x_2)]^{\nu_1}
[W^\dagger_2(x_1)]^{\nu_2}
f_j (\vec{\nu})
\label{eq:Uj}
,\quad
\end{eqnarray}
where the coordinate dependence is made explicit through the appearance of Wilson loops. 
One-dimensional integral representations for these potentials can be found from the expressions collected in 
Appendix~\ref{s:A}. 
These residual potentials isolate the non-trivial coordinate dependence of the neutron effective action. 
Utilizing the long Euclidean time limit of the neutron two-point correlation function, 
one will arrive at contributions from the lowest eigenstate of the Hamiltonian
\begin{equation}
H
=
- 
\frac{\vec{\nabla}_\perp^2}{2 M}
+ 
U_0 (\vec{x}_\perp)
+ 
e B \sigma_3 \,
U_1 ( \vec{x}_\perp)
\label{eq:ham}
,\end{equation}
which is the non-trivial part of the neutron effective action. 
The corresponding lowest energy eigenvalue,
$\mathcal{E}_0$, 
we write in terms of spin-independent and spin-dependent contributions
\begin{equation}
\mathcal{E}_0 
= \lambda_{0,0} (L) + e B \sigma_3 \, \lambda_{1,0} (L)
.\end{equation} 
These eigenvalues lead to additional finite volume corrections to the neutron energy, 
and hence to the magnetic moment and magnetic polarizability. 
We will argue, 
however, 
that finite volume effects from coordinate-dependent potentials
can be treated in perturbation theory,
and are suppressed by a power of 
$p$
because they require neutron recoil. 
As a result, 
$\Delta E_0$
and
$\Delta E_1$
can be identified as the finite volume corrections to the neutron energy up to recoil corrections.

To argue that the effect of residual potentials on the neutron energy is suppressed,
we scrutinize the power counting of the effective Hamiltonian in Eq.~\eqref{eq:ham}. 
The kinetic energy operator represents an 
$\cO(p^2)$
contribution that ordinarily can be dropped by considering a neutron at rest. 
The coordinate dependence of the residual potentials, 
however, 
forces us to reconsider the Fourier modes of the neutron. 
The spin-independent residual potential 
$U_0(\vec{x}_\perp)$
scales as 
$\cO(p^3)$
while the spin-dependent residual potential 
$U_1(\vec{x}_\perp)$
scales as 
$\cO(p)$. 
The effect of both potentials on the neutron energy is thus at
$\cO(p^3)$. 
Consequently we can consider the unperturbed neutron states as Fourier modes at 
$\cO(p^2)$
and address the effect of the residual potentials in perturbation theory. 
Because the unperturbed ground state is a neutron at rest, 
$\vec{P} = \vec{0}$, 
we have the vanishing of the unperturbed energy eigenvalue of the Hamiltonian, 
$\mathcal{E}_0^{(0)} = 0$. 
The leading correction to the non-degenerate ground-state energy, 
$\mathcal{E}_0^{(1)}$,  
scales as
$\cO(p^3)$
and is determined from the matrix elements
\begin{eqnarray}
\lambda_{j,0}^{(1)} 
&=&
\langle \vec{P} = \vec{0} \, | U_j | \vec{P} = \vec{0} \, \rangle
\notag \\
&=&
\sum_{\vec{\nu} - \{ \vec{0},  \vec{0}_\perp \} }
\int_0^L d\vec{x}_\perp
e^{ 2 \pi i N_\Phi ( \nu_1 x_2 - \nu_2 x_1 ) / L}
f_j (\vec{\nu})
=
0,\notag \\
\end{eqnarray}
where in the second line we have used the fact that images with 
$\vec{\nu}_\perp = \vec{0}_\perp$
have been explicitly removed from the summation over winding number. 
The vanishing of the ground-state matrix elements has a simple physical interpretation. 
In the Fourier basis,  
Wilson loops produced by the 
$\vec{\nu}$-th image 
insert momentum
\begin{equation}
\vec{q}_\perp = \frac{2 \pi N_\Phi}{L} ( - \nu_2, \nu_1),
\end{equation}
into the matrix elements of residual potentials. 
Thus all forward matrix elements of the perturbation vanish. 
One must look to degenerate Fourier modes to arrive at a non-vanishing 
first-order perturbation that scales as 
$\cO(p^3)$
in our power counting. 
As a consequence, 
the ground state energy 
$\mathcal{E}_0$
has finite volume corrections arising from the residual coordinate-dependent potentials at 
$\cO(p^4)$, 
which is beyond the order we are working.


%
\begin{figure}
\begin{flushleft}
\epsfig{file=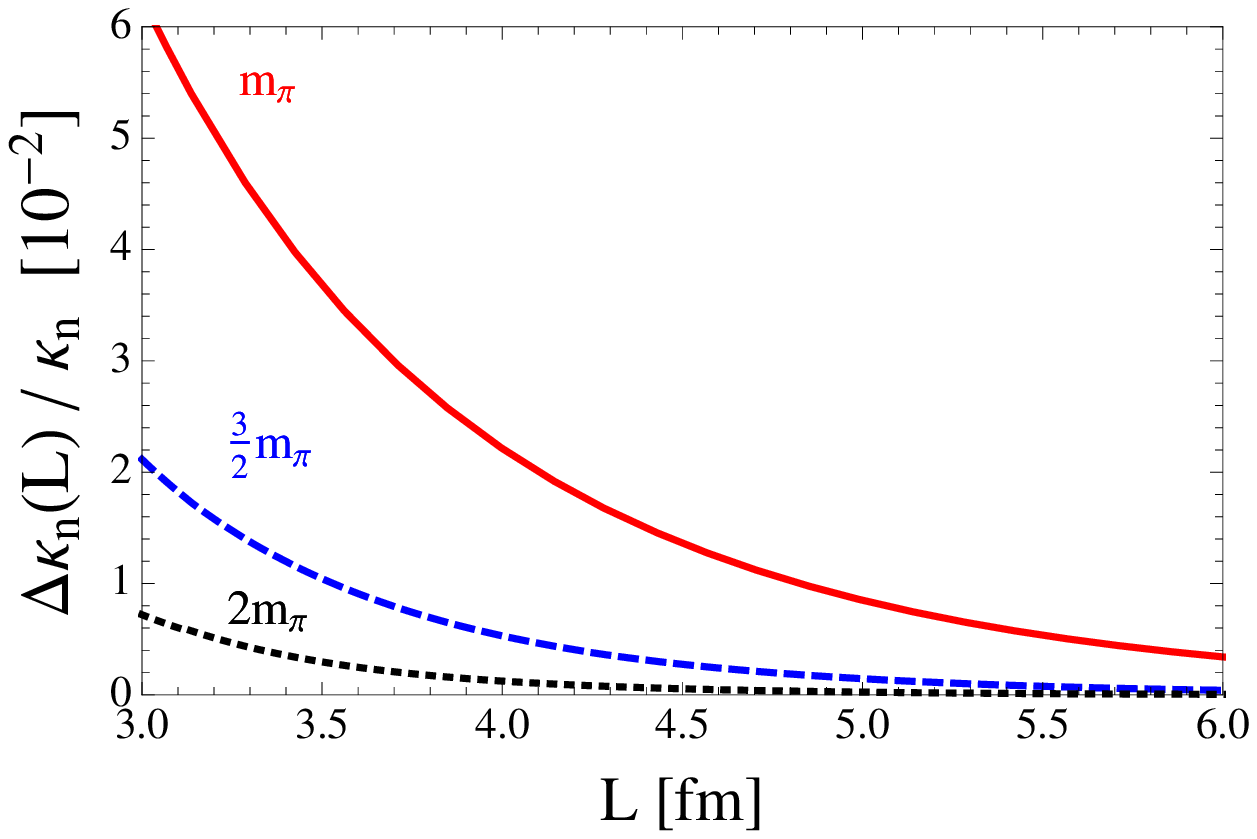,width=8.25cm}
\\
\smallskip
\smallskip
\epsfig{file=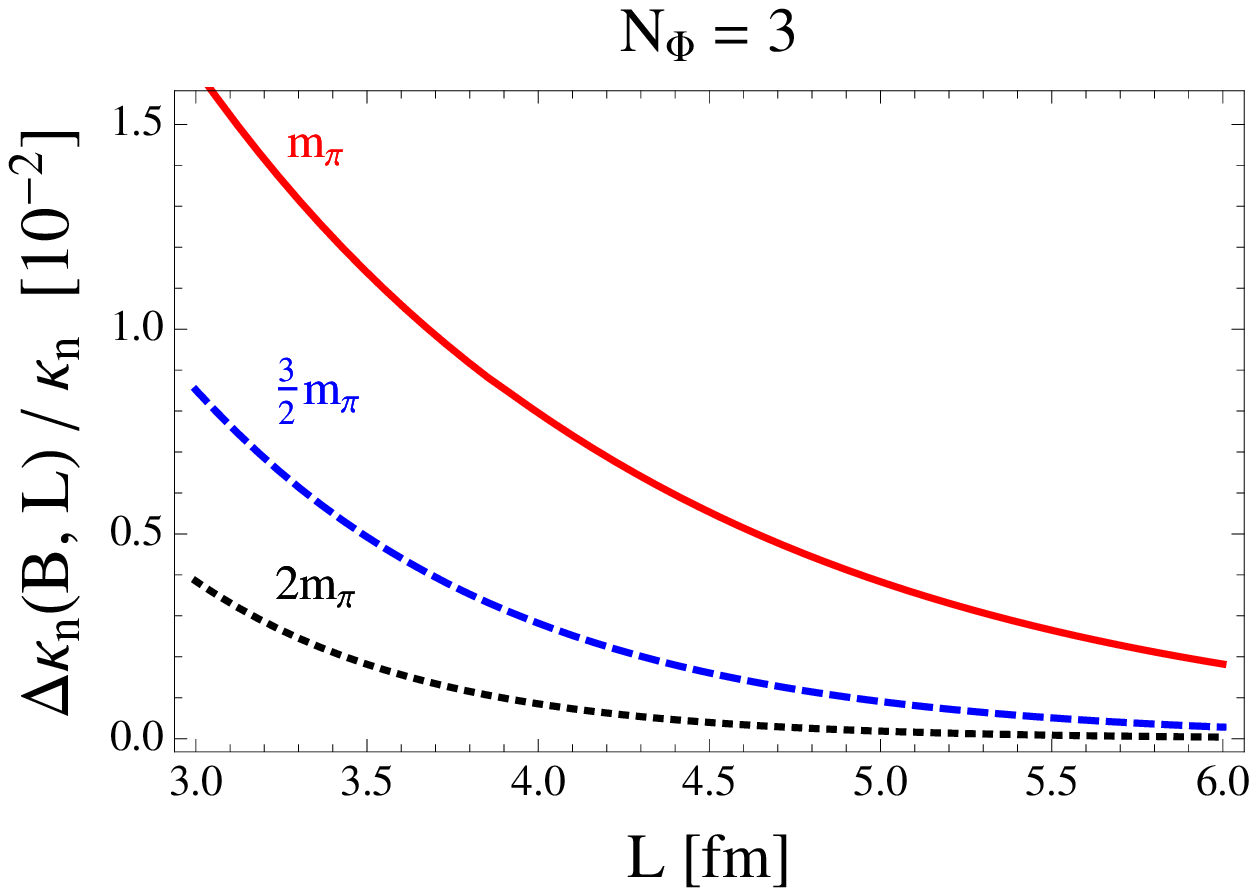,width=8.1cm}
\caption{\label{f:kappaNP} 
Finite volume effect on the ground-state neutron Zeeman splitting computed in weak and strong magnetic fields. 
Plotted versus 
$L$
are the finite volume corrections 
$\Delta \kappa_n (L)$
and the strong-field generalization
$\Delta \kappa_n (B,L)$. 
These corrections are compared to the physical neutron magnetic moment, 
$\kappa_n = - 1.91 \, \texttt{[} \mu_N \texttt{]}$. 
Results are shown using three values of the pion mass, 
and include the effects of proton-pion and delta-pion intermediate states. 
The strong-field results are shown for a flux quantum of 
$N_\Phi = 3$. 
                }%
                \end{flushleft}
        \end{figure}
%



%
\begin{figure}
\begin{flushleft}
\epsfig{file=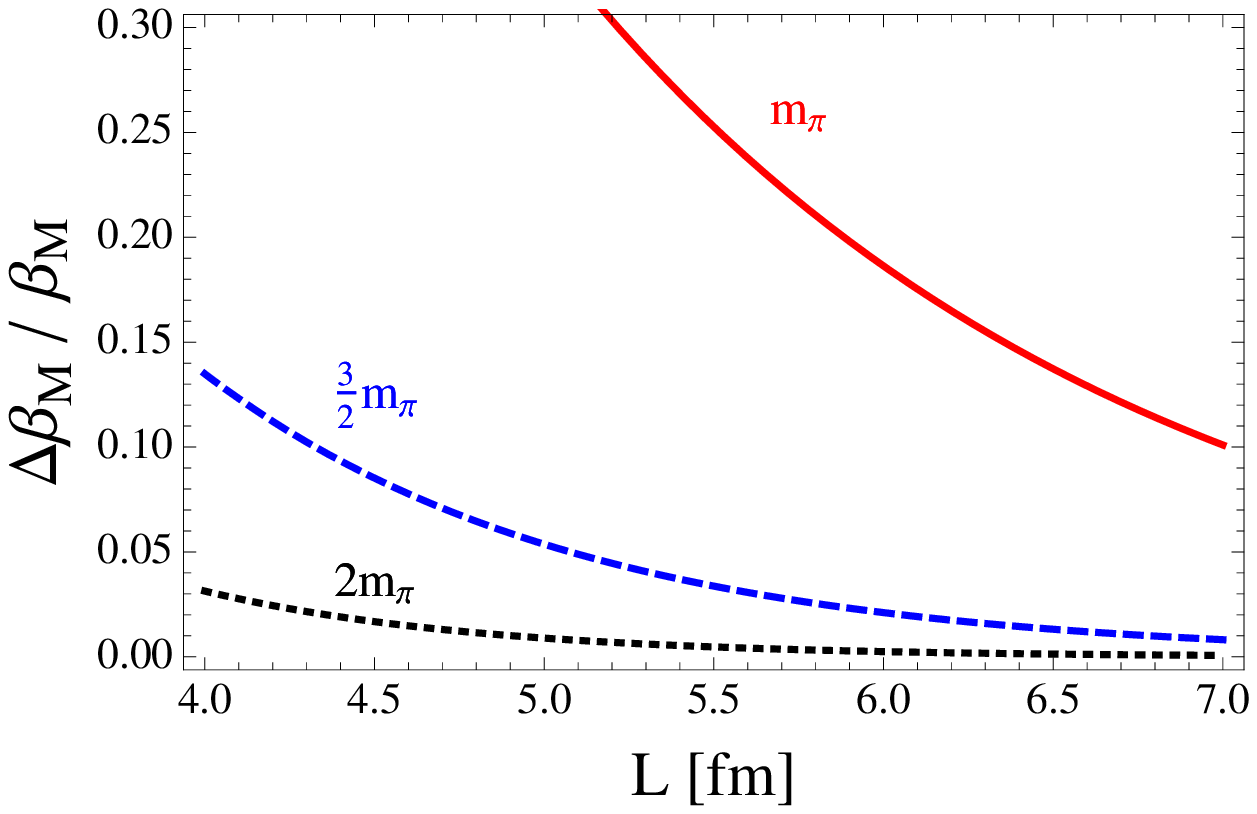,width=8.25cm}
\\
\smallskip
\smallskip
\epsfig{file=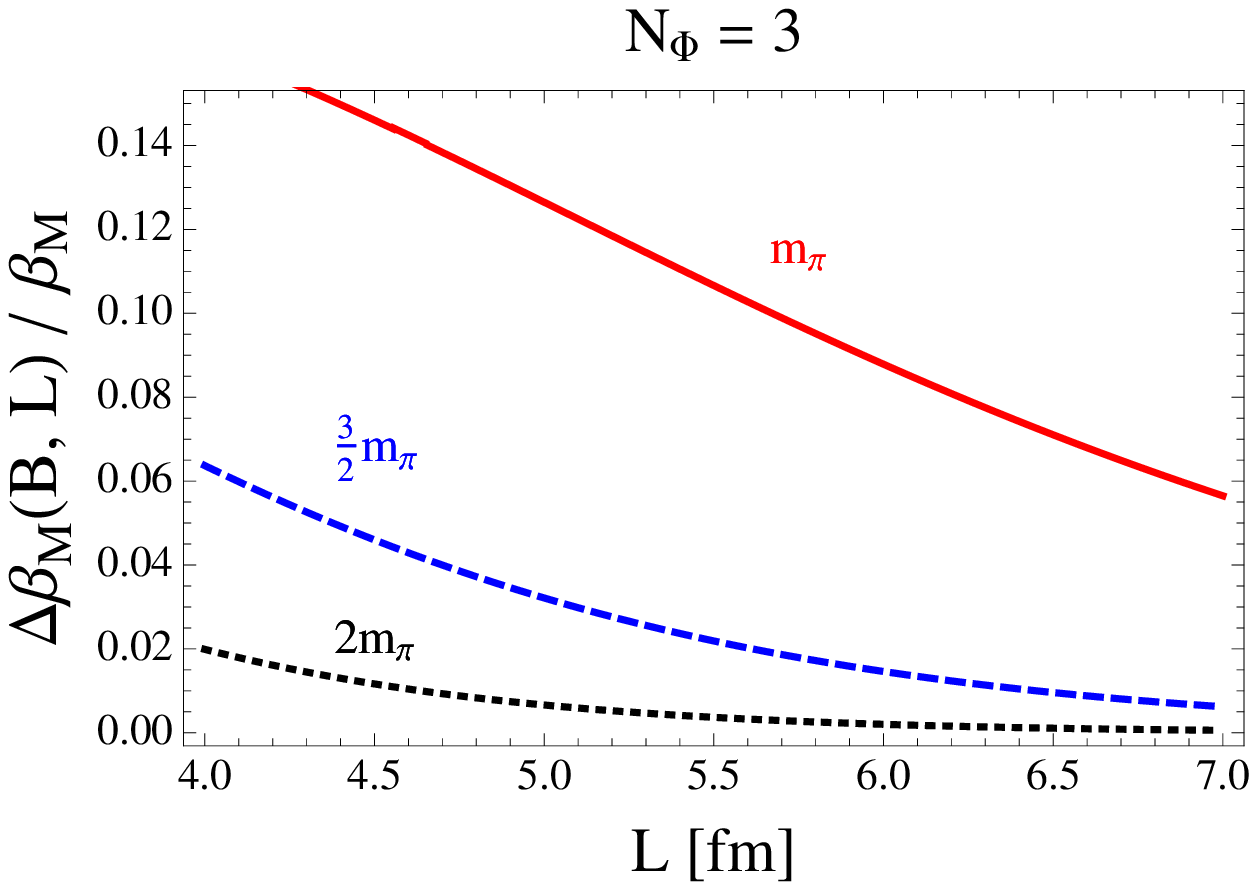,width=8.25cm}
\caption{\label{f:beta} 
Finite volume effect on the spin-independent ground-state neutron energy. 
Plotted versus the lattice size
$L$
are the finite volume corrections
$\Delta \beta_M(L)$
and the strong-field generalization 
$\Delta \beta_M (B, L)$. 
These finite-size effects are compared to
$\beta_M 
\equiv
\frac{e^2}{4 \pi} 
\frac{g_A^2}
{96 \pi f^2 m_\pi} 
= 
1.25 \times 10^{-4} \, \texttt{[fm]}^3$, 
which we take as an estimate of the physical value. 
Results are shown using three values of the pion mass. 
Included are contributions from both proton-pion and delta-pion intermediate states. 
The strong-field results are evaluated for a flux quantum 
$N_\Phi = 3$. 
               }%
                \end{flushleft}
        \end{figure}
%


To assess the size of finite volume corrections on the ground-state neutron, 
we first consider the weak-field regime, 
$|e B| / m_\pi^2 \ll 1$. 
In this regime, 
we can evaluate the 
$\cO(p^3)$
finite volume effect on the spin-dependent neutron energy  
$\Delta E_1 (L)$
appearing in Eq.~\eqref{eq:E1}
at vanishing magnetic field. 
Notice all of the topological effects that complicate taking the weak-field limit reside in the coordinate-dependent potentials that affect the ground-state energy at 
$\cO(p^4)$. 
The leading finite-volume correction to the magnetic moment can easily be deduced by performing the sum over winding number. 
We arrive at the expression
\begin{equation}
\Delta 
\kappa_n (L)
=
\frac{g_A^2 M}{2 \pi f^2 L}
\log ( 1 - e^{ - m_\pi L} )
\label{eq:kappa3}
.\end{equation}
Focusing on the spin-independent contribution, 
we can write
\begin{equation}
\Delta E_0 (L) 
=
- \frac{1}{2} 4 \pi \,
\Delta \beta_M(L) 
\, B^2  
+ 
\cdots
,\end{equation}
in the weak-field regime. 
This simple expansion is valid due to the lack of topological contributions at this order, 
and enables us to identify the finite volume correction to the magnetic polarizability as
\begin{equation}
\Delta 
\beta_M (L)
=
\frac{e^2}{4 \pi}
\frac{g_A^2 L}{48 \pi f^2}
\frac{e^{ - m_\pi L}}{( 1 - e^{ - m_\pi L} )^2}
\left[
1 + \frac{ 1 - e^{ - m_\pi L}}{m_\pi L}
\right]
\label{eq:beta3}
.\end{equation}


%
\begin{figure}
\begin{flushleft}
\epsfig{file=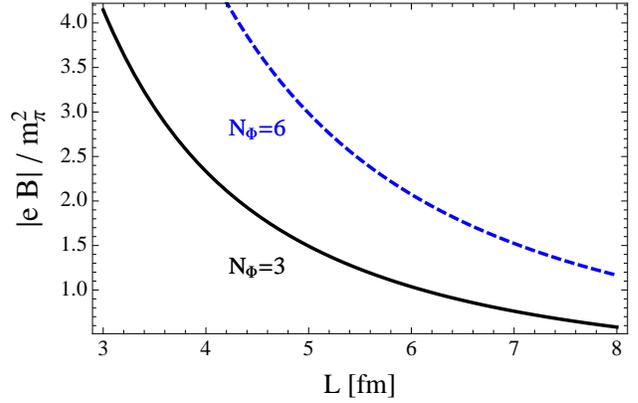,width=8.25cm}
\caption{\label{f:xi} 
Plot of the weak-field expansion parameter 
$| \xi | = | e B | / m_\pi^2$
as a function of the lattice size
$L$. 
The magnetic field is quantized according to 
$e B  L^2  = 2 \pi N_\Phi$, 
and we use the physical value of the pion mass.
The expansion parameter is shown for the two lowest values of the flux quantum for QCD. 
                }%
                \end{flushleft}
        \end{figure}
%


The size of finite volume corrections to the neutron magnetic moment and magnetic polarizability is explored in 
Figs.~\ref{f:kappaNP} and \ref{f:beta}. 
Included in the plot are the effects from proton-pion intermediate states given in this section, 
and effects from delta-pion intermediate states.  
The computation including delta degrees of freedom has been relegated for simplicity to Appendix~\ref{s:D}. 
Specifically relevant are the finite volume results shown in 
Eqs.~\eqref{eq:FVmomentNDelta} and \eqref{eq:betaFVDelta}. 
A practical consideration concerns the applicability of weak-field results to lattice QCD computations. 
While lattice volumes have increased in physical units leading to smaller values of magnetic fields, 
the lattice pion masses have decreased towards the physical point.
As a result, 
the expansion parameter 
$ e B / m_\pi^2$
has roughly remained the same. 
The weak-field results determined above are thus contrasted with the more realistic strong-field results, 
for which 
$| e B | / m_\pi^2 \sim 1$. 
The latter results are determined from the energy shifts shown in
Eqs.~\eqref{eq:E0} and \eqref{eq:E1},
via the relations
\begin{eqnarray}
\Delta \kappa_n (B, L)
&=&
2 M \Delta E_1 (L),
\end{eqnarray}
and
\begin{eqnarray}
\Delta \beta_M (B, L) 
&=&
\Delta E_0 (L) \Big/ \left( - \frac{1}{2} 4 \pi B^2 \right)
.\end{eqnarray}
These strong-field generalizations of finite-volume corrections to the magnetic moment and magnetic polarizability properly reduce in the weak-field limit. 
In order to investigate the finite volume corrections in the strong-field regime, 
we evaluate the effects using magnetic fields available on the torus, 
$e B L^2 = 2 \pi N_\Phi$.  
For reference, 
the size of the weak-field expansion parameter, 
$\xi = e B / m_\pi^2$,
is shown as a function of lattice size
$L$
in Fig.~\ref{f:xi}. 
Rather large volumes are needed for strong-field results to apply at the physical pion mass. 
For this reason, 
we plot the strong-field finite volume effects in 
Figs.~\ref{f:kappaNP} and \ref{f:beta} choosing the smallest possible magnetic field for QCD, 
which requires
$N_\Phi = 3$.
Doubling the pion mass of course allows one to utilize magnetic fields with up to four times the strength. 
Despite the size of magnetic fields allowed on small volumes, 
the finite volume corrections to the magnetic moment remain considerably small. 
For this reason, 
we have optimistically plotted the results for the magnetic moment in 
Fig.~\ref{f:kappaNP} 
down to lattice sizes of 
$L = 3 \, \texttt{fm}$. 
On the other hand, 
larger finite-size effects are seen for the magnetic polarizability. 
As a result, 
we have opted to be more conservative about the minimum-size lattice for which our results for the spin-independent energy apply, 
and have limited the plots for the polarizability 
in 
Fig.~\ref{f:beta} 
to sizes
$L > 4 \, \texttt{fm}$. 
For both spin-independent and spin-dependent finite volume corrections, 
we see a sizeable reduction in finite volume effects when strong-field results are utilized. 
This signals a breakdown of the weak-field expansion for the lattice sizes and pion masses considered. 
The need to treat the magnetic field non-perturbatively is further evidenced by Fig.~\ref{f:xi}.

\subsubsection{Asymptotic Volume Limit}

We entertain a final limit characterized by the possibility to obtain analytic results for the neutron wave functions.
This limit is that of asymptotically large volumes. 
In the large volume limit, 
we can restrict our attention to sectors of minimal winding number, 
with neglected terms being exponentially suppressed.
In this limit, 
the magnetic field becomes small for fixed magnetic flux quantum,  
$e B L^2 = 2 \pi N_\Phi$,
so we can safely assume that the weak-field limit, 
$| e B | / m_\pi^2 \ll 1$,
will additionally be satisfied.

For the neutron effective action, 
we focus exclusively on the coordinate dependence, 
and take into account contributions from the sectors with winding number
$\vec{\nu} = ( \pm 1, 0, 0)$
and 
$\vec{\nu} = ( 0, \pm 1, 0)$. 
In the weak-field limit, 
only terms from the spin-independent potential in 
Eq.~\eqref{eq:V0}
survive at leading order.
These images produce the potential
\begin{eqnarray}
\cW (\vec{x}_\perp)
=
\cW
\left[
\cos ( e B L x_1) + \cos ( e B L x_2)
\right]
,\end{eqnarray}
with the overall strength of the potential set by the parameter
\begin{eqnarray}
\cW
&=& 
\frac{g_A^2 m_\pi^2}{4 \pi f^2 L} e^{ - m_\pi L}
,\end{eqnarray} 
which scales as 
$\cO(p^3)$
in the power counting. 
Further images yield contributions that are exponentially suppressed. 
As a consequence, 
the transverse motion of the neutron is determined by the effective Hamiltonian, 
$H_\text{eff} = - \frac{\vec{\nabla}_\perp^2}{2 M} + \cW ( \vec{x}_\perp)$, 
and the solutions factorize into products of solutions to the one-dimensional Hamiltonian, 
\begin{equation}
H
= 
- \frac{1}{2 M} \frac{d^2}{d x^2}
+ 
\cW \cos ( e B L x)
\label{eq:M}.\end{equation}
The corresponding Schr\"odinger equation that determines the energy eigenvalues is the Mathieu differential equation, 
which has the canonical form
\begin{equation}
\left[\frac{d^2}{dv^2} + \alpha - 2 q \cos ( 2 v ) \right] \psi(v) = 0
.\end{equation} 
Translating the parameters of our problem into Mathieu's equation, 
we have the transverse coordinates given by
$\vec{v}_\perp = \pi N_\Phi \vec{x}_\perp / L$,
and
$q = M L^2 \cW / \pi^2 N_\Phi^2$.

Given the periodicity of neutron interpolating operators in lattice QCD, 
and the form of the potential in
Eq.~\eqref{eq:M}, 
we must seek solutions that are 
$\mathbb{Z}_{N_\Phi}$ 
translationally invariant, 
which corresponds to 
$\pi$-periodicity in the coordinates 
$\vec{v}_\perp$. 
The ground-state solution is thus characterized by the wave function
\begin{equation}
\psi_0 (\vec{x}_\perp )
=
N 
\texttt{ce}_0 \left( v_1,  q \right)
\texttt{ce}_0 \left( v_2,  q \right)
,\end{equation}
where 
$\texttt{ce}_r (v, q)$
denotes a Mathieu function, 
and 
$N$
is an overall normalization factor. 
The corresponding ground-state energy 
is given by
\begin{equation}
E(L)
= 
\frac{\pi^2 N_\Phi^2}{M L^2} \, 
\texttt{a}_0 \left( q \right)
,\end{equation}
where
$\texttt{a}_r (q)$
denotes a Mathieu characteristic. 
Using the series expansion, 
$\texttt{a}_0(q) = - \frac{1}{2} q^2 + \cO(q^4)$, 
we see the energy eigenvalue can be written as
\begin{equation}
E(L) 
= 
- \frac{1}{2} M L^2 \frac{\cW^2}{\pi^2 N_\Phi^2} 
+ 
\cdots
.\end{equation}
The leading term in this expansion scales as 
$\cO(p^4)$, 
which is consistent with our general analysis above.

\subsection{Conclusion}

Above we consider chiral dynamics of the neutron in a strong external magnetic field, 
including a derivation of finite volume corrections relevant for uniform magnetic fields on a torus. 
In order to perform these computations, 
a number of theoretical ingredients are established. 
Magnetic periodicity is enforced on charged scalar Green's functions, 
resulting in the coordinate-space propagator in Eq.~\eqref{eq:GFV}. 
The features of this propagator allow us to anticipate many general properties of the finite volume computation. 
In particular, 
Wilson loops lead to the breaking of translational invariance down to the discrete
$\mathbb{Z}_{N_\Phi}$
magnetic translation group. 
These Wilson loops must accompany the winding of charged particles around the torus due to gauge invariance. 
Furthermore, 
a phase factor in the infinite volume propagator, 
Eq.~\eqref{eq:Bprop}, 
underlies the appearance of the Aharonov-Bohm effect in our later computations.  
We are careful to show that the static charged fermion propagator is unaffected by magnetic periodic images.

The charged scalar and static fermion propagators are utilized to compute chiral corrections to the correlation functions of the neutron. 
This is accomplished for the first time using a combination of heavy nucleon chiral perturbation theory, 
$p$-regime chiral perturbation theory for finite volume effects, 
and strong-field chiral counting to handle non-perturbative effects from the magnetic field. 
The finite volume computations are performed using a direct coordinate-space method that circumvents redundant application of the Poisson formula. 
The coordinate-space computation is applied to a simple example to recover the familiar finite volume corrections to the nucleon mass. 
After this demonstration,
computations in magnetic fields are performed.

In infinite volume, 
the dependence of the neutron energy on the magnetic field is considered for strong magnetic fields satisfying the condition
$| e B | / m_\pi^2 \gtrsim 1$. 
In Fig.~\ref{f:E0}, 
we show the spin-independent neutron energy shift in a strong magnetic field. 
While the perturbative expansion in powers of the magnetic field breaks down, 
our results demonstrate that a model incorporating the leading perturbative effect from the magnetic polarizability may still capture the non-perturbative behavior, 
however, 
the polarizability extracted from a perturbative model fit will differ from the true value. 
Comparison of perturbative and non-perturbative magnetic field fits will be necessary to assess this effect. 
A very similar observation is made about the neutron Zeeman splitting shown in Fig.~\ref{f:E1}, 
which shows surprisingly linear behavior with respect to the magnetic field well beyond the perturbative magnetic field regime. 
In this case, 
higher-order chiral corrections are needed from the effective theory to verify the behavior at the largest magnetic fields 
shown in the plot. 
It would be quite fortuitous if this behavior persists.

Finite volume corrections are derived for the neutron in the form of terms appearing in the neutron effective action, 
which arises after integrating out charged pions. 
The proton-pion (and delta-pion) intermediate states lead to coordinate dependence of the neutron effective action. 
Winding of charged pions around the torus must be accompanied by Wilson loops, 
and these gauge invariant quantities reflect the non-trivial holonomies of the torus. 
A topological phase can be acquired from pions winding in the plane transverse to the field, 
which is a manifestation of the Aharonov-Bohm effect. 
Exposing these features requires non-perturbative treatment of the magnetic field at finite volume. 
Consequently these features are not susceptible to perturbative expansion in powers of the magnetic field except in rather extreme limits. 
Considering such limits, 
we obtain formulae for the finite volume effects on the magnetic moment and magnetic polarizability. 
The former result agrees with that derived using twisted boundary conditions in the limit of a vanishing twist angle. 
We stress that the magnetic moments and magnetic polarizabilities do not have unique definitions on a torus. 
As such they are subject to finite volume effects that depend on the method employed in their determination. 
In the case of three- and four-point function computations, 
the breaking of rotational invariance is a complication that has often been overlooked in finite volume.
In the external field problem, 
topological effects from holonomies and Aharonov-Bohm phases are the new features at finite volume.

To evaluate the effect of finite volume on the neutron correlation function, 
we are led beyond the static approximation. 
The coordinate dependence of the neutron potential requires that the neutron kinetic term be retained. 
While Fourier modes no longer exactly describe the propagation of the neutron in finite volume, 
the finite volume corrections are suppressed relative to the kinetic term. 
Standard Rayleigh-Schr\"odinger perturbation theory can be employed to determine the effect on neutron energy levels. 
Focusing on the ground-state neutron energy, 
we determine the finite volume corrections, 
and uncover that the coordinate-dependent potentials do not affect results at the leading order of perturbation theory. 
As a result, 
the leading finite volume corrections for neutron magnetic moments and polarizabilities originate solely from pions winding in the direction of the magnetic field. 
Physically we might expect suppression of winding transverse to the field direction because the magnetic field provides a  confining length scale, 
$\sim | e B |^{-1/2}$,
that is characteristic of Landau levels.  
In the evaluation of finite volume effects, 
Figs.~\ref{f:kappaNP} and \ref{f:beta}, 
we consider results for both the weak-field limit and for strong magnetic fields. 
The latter are readily encountered on present-day lattices, 
and we see non-perturbative magnetic field effects decrease the size of finite volume corrections.

Various further investigations have been suggested throughout. 
Here we describe a few that have yet to be mentioned. 
Central to our discussion has been the static approximation for the neutron. 
As the neutron effective potential on a torus is coordinate dependent, 
the static approximation must be reconsidered.
Away from the static limit, 
strong magnetic fields in infinite volume additionally probe hadron structure, 
and allow for further coordinate dependence in neutron correlation functions beyond what is considered here. 
Such dependence can be investigated in the context of chiral perturbation theory with strong-field power counting.   
A study of this coordinate dependence in conjunction with the coordinate dependence exposed above may help in the construction of better neutron interpolating operators for lattice computations. 
Finally our computation has been limited to correlation functions of the neutron only in the interest of simplicity. 
Proton correlation functions can be treated in an entirely similar manner to that developed in this work. 
In the case of the proton, 
one must understand the modifications to Landau levels that arise both from the torus, 
and from charged pions winding around the torus. 
Nevertheless, 
we imagine that the methodology of strong-field chiral perturbation theory explored here will be of use to 
lattice QCD computations of hadrons in electromagnetic fields.

\begin{acknowledgments}
Work supported in part by a joint City College of New York--RIKEN BNL Research Center fellowship, 
a grant from the Professional Staff Congress of the CUNY, 
the Alfred P.~Sloan foundation through a CUNY-JFRASE award, 
and by the U.~S.~National Science Foundation, 
under Grant No.~PHY$12$-$05778$.
\end{acknowledgments}

\appendix

\begin{widetext}

\section{Including the Delta Resonance}
\label{s:D}

Of the low-lying baryon resonances, 
the nearby
delta resonance gives rise to important virtual corrections to nucleon observables. 
The delta-nucleon mass splitting, 
$\Delta = M_\Delta - M_N$,
introduces a new dimensionful parameter, 
and it is common to treat it as 
$\Delta / M \sim p$
in the power counting, 
see~\cite{Hemmert:1997ye}.  
The 
$\cO(p)$
chiral 
Lagrange density describing the 
delta resonance
and its interactions with the nucleon is given by
\begin{eqnarray}
\cL
&=&
\ol T_\mu \left( D_4 + \Delta \right) T_\mu 
+ 
g_{\Delta N}
\left[
\ol T_\mu \cA_\mu N 
+
\ol N \cA_\mu T_\mu 
\right],
\label{eq:T}
\end{eqnarray}
where the delta-nucleon axial coupling is 
$g_{\Delta N} \sim 1.5$. 
The magnetic polarizability receives a contribution from the nucleon-delta magnetic dipole transition operator. 
As the leading contribution is from a tree-level diagram with intermediate-state delta, 
there is no volume correction associated with it.

To derive the leading finite volume corrections to the neutron two-point function, 
we evaluate the four sunset diagrams shown in Fig.~\ref{f:sunset}, 
where the intermediate-state baryon is now taken to be a 
delta resonance. 
From the Lagrange density in 
Eq.~\eqref{eq:T}, 
we see the static delta propagator is given by
\begin{equation}
[D_{FV} (x',x)]_{ij}
= 
\delta_{L}^{(3)}
(\vec{x} \, ' - \vec{x} \, )
\theta( x'_4 - x_4)
e^{ - \Delta ( x'_4 - x_4)}
\cP_{ij}
\label{eq:D1.5}
,\end{equation}
where the 
spin-$\frac{3}{2}$ 
projection matrix has the form 
$\cP_{ij} = \frac{2}{3} \delta_{ij} - \frac{i}{3} \epsilon_{ijk} \sigma_k$. 
The proper-time formulation makes accounting for 
delta-resonance contributions straightforward. 
By virtue of the propagator,
Eq.~\eqref{eq:D1.5},
the effect of the mass splitting 
$\Delta$
is to alter the relative-time integration used to project onto 
$P_4 = 0$, 
where we encounter the integrals
\begin{eqnarray}
\int_0^\infty
dT \, 
e^{ - \frac{T^2}{4 s} - \Delta \, T}
&=&
\sqrt{\pi s} \,
e^{\Delta^2 s}
\Erfc
\left( \Delta \sqrt{s} \right), 
\qquad
\int_0^\infty 
dT \, 
\frac{T}{2s}
e^{ - \frac{T^2}{4 s} - \Delta \, T}
=
1 - 
\Delta
\sqrt{\pi s} \,
e^{\Delta^2 s}
\Erfc
\left( \Delta \sqrt{s} \right)
.\end{eqnarray} 
Here 
$\Erfc(x)$
is the complement of the standard error function, 
for which 
$\Erfc(0) = 1$, 
and the second integral arises from utilizing an integration by parts on the relative-time integral. 
Such contribution only occurs for the spin-independent part of the sunset diagrams.

Combining the result of the relative-time integration with the spin and flavor algebra, 
we have the 
delta-resonance contributions to the infinite volume quantities:
the spin-independent energy
\begin{eqnarray}
E_0^{(\Delta)}
&=&
\frac{8}{9}
\frac{g_{\D N}^2 \sqrt{\pi}}{(4 \pi f)^2}
\int_0^\infty \frac{ds}{s^{3/2}}
\left(
\frac{e B s}{\sinh e B s}
-
1
\right)
e^{ - m_\pi^2 s} 
\left[
\frac{\Delta}{\sqrt{\pi s}}
+
\left( m_\pi^2 - \Delta^2 \right)
e^{ \Delta^2 s} 
\Erfc
\left( \Delta \sqrt{s} \right)
\right], 
\end{eqnarray}
so that the sum 
$E_0 + E_0^{(\D)}$
takes into account intermediate-state nucleons and deltas, 
with 
$E_0$
is given in 
Eq.~\eqref{eq:strong0};
and the spin-dependent energy
\begin{eqnarray}
E_1^{(\Delta)}
&=&
\frac{2}{9}
\frac{g_{\D N}^2 \sqrt{\pi}}{(4 \pi f)^2}
\int_0^\infty \frac{ds}{s^{3/2}}
\left(
\frac{e B s}{\sinh e B s}
-
1
\right)
e^{ - ( m_\pi^2 - \Delta^2) s} 
\Erfc
\left( \Delta \sqrt{s} \right)
,\end{eqnarray}
so that similarly 
$E_1 + E_1^{(\D)}$
takes into account both intermediate-state baryons, 
with 
$E_1$
given in 
Eq.~\eqref{eq:strong1}. 
The differing numerical constants for spin-independent and spin-dependent contributions stem from the underlying 
spin $\times$ flavor $\times$ charge factors of intermediate states. 
In each case, 
we have subtracted the divergent zero-field results. 
The dimensionally regulated zero-field results produce the charged pion loop corrections to the 
neutron mass and magnetic moment, 
in the spin-independent and spin-dependent cases, respectively. 
Furthermore, 
the finite
$\cO(B^2)$
term in the spin-independent energy leads to the correct contribution to the neutron's magnetic polarizability, 
originally determined in~\cite{Hemmert:1996rw}.

The finite volume corrections with intermediate-state delta-resonance contributions can be similarly found. 
We can write their contribution in the form 
\begin{eqnarray}
V_0^{(\D)} (\vec{x}_\perp)
&=&
\frac{8}{9}
\frac{g_{\D N}^2 \sqrt{\pi}}{(4 \pi f)^2}
\int_0^\infty \frac{ds}{s^{3/2}}
\frac{e B s}{\sinh e B s}
e^{ - m_\pi^2  s} 
\Bigg\{
\left[
\frac{\Delta}{\sqrt{\pi s}}
+
\left( m_\pi^2 - \Delta^2 \right)
e^{ \Delta^2 s} 
\Erfc
\left( \Delta \sqrt{s} \right)
\right]
\mathcal{U}^{(A)} (s, B, \vec{x}_\perp)
\notag \\
&& \phantom{spacingspacing}
+ 
\frac{1}{2} ( e B L)^2
e^{ \Delta^2 s} 
\Erfc
\left( \Delta \sqrt{s} \right)
\mathcal{U}^{(B)} (s, B, \vec{x}_\perp)
\Bigg\}
\label{eq:V0Delta}
,\end{eqnarray}
for the spin-independent part, 
and 
\begin{eqnarray}
V_1^{(\D)} (\vec{x}_\perp)
&=&
\frac{2}{9}
\frac{g_{\D N}^2 \sqrt{\pi}}{(4 \pi f)^2}
\int_0^\infty \frac{ds}{s^{3/2}}
\frac{e B s}{\sinh e B s}
e^{ - ( m_\pi^2 - \Delta^2) s} 
\Erfc
\left( \Delta \sqrt{s} \right)
\mathcal{U}_1 (s, B, \vec{x}_\perp)
,\label{eq:V1Delta}
\end{eqnarray}
for the spin-dependent part. 
The functions 
$\mathcal{U}^{(A)} (s, B, \vec{x}_\perp)$, 
$\mathcal{U}^{(B)} (s, B, \vec{x}_\perp)$, 
and
$\mathcal{U}_1 (s, B, \vec{x}_\perp)$
are given below in 
Eqs.~\eqref{eq:spindepU1}--\eqref{eq:UB}.

In the main text, 
we present expressions for finite volume corrections to the neutron magnetic moment and magnetic polarizability arising from proton-pion intermediate states. 
Here we quote the analogous expressions taking into account contributions from delta-pion intermediate states. 
For the magnetic moment defined by taking extreme limits, 
we have
\begin{eqnarray}
\Delta \kappa_n^{(\Delta)} (L)
=
\frac{4}{9}
\frac{g_{\D N}^2 M \sqrt{\pi}}{(4 \pi f)^2}
\int_0^\infty \frac{ds}{s^{3/2}}
e^{ - ( m_\pi^2 - \Delta^2) s} 
\Erfc
\left( \Delta \sqrt{s} \right)
\mathcal{U}_1 (s, 0, 0)
\label{eq:FVmomentNDelta}
,\end{eqnarray}
where
$\mathcal{U}_1 (s, 0, 0)$
is defined without dependence on the magnetic flux quantum 
$N_\Phi$,
and appears in Eq.~\eqref{eq:U0s}. 
For the magnetic polarizability defined by taking extreme limits, 
we have
\begin{eqnarray}
\Delta \beta_M^{(\Delta)} (L)
&=&
\frac{e^2}{4 \pi}
\frac{8}{9}
\frac{g_{\D N}^2 \sqrt{\pi}}{(4 \pi f)^2}
\int_0^\infty \frac{ds}{s^{3/2}}
e^{ - m_\pi^2  s} 
\Bigg\{
\frac{1}{3} s^2
\left[
\frac{\Delta}{\sqrt{\pi s}}
+
\left( m_\pi^2 - \Delta^2 \right)
e^{ \Delta^2 s} 
\Erfc
\left( \Delta \sqrt{s} \right)
\right]
\mathcal{U}^{(A)} (s, 0, 0)
\notag \\
&& \phantom{spacingspace}
+
\frac{1}{6} s L^2
\left[
\frac{\Delta}{\sqrt{\pi s}}
+
\left( m_\pi^2 - \Delta^2 - \frac{6}{s} \right)
e^{ \Delta^2 s} 
\Erfc
\left( \Delta \sqrt{s} \right)
\right]
\mathcal{U}^{(B)} (s, 0, 0)
\Bigg\}
\label{eq:betaFVDelta}
,\end{eqnarray}
with expressions for the functions 
$\mathcal{U}^{(A)} (s, 0, 0)$
and
$\mathcal{U}^{(B)} (s, 0, 0)$
given in 
Eq.~\eqref{eq:UAB0s}.

In strong-field power counting, 
we obtain simple expressions for spin-independent and spin-dependent finite volume effects on the ground-state 
neutron. 
The simple expressions arise because non-vanishing corrections from coordinate-dependent potentials in the neutron effective action occur at 
$\cO(p^4)$. 
The argument applies equally for contributions arising from delta-pion intermediate states. 
For strong-field corrections to the neutron ground-state energy in finite volume, 
we thus have the results
\begin{eqnarray}
\Delta \kappa_n^{(\Delta)} (B, L)
=
-
\frac{4}{9}
\frac{g_{\D N}^2 M \sqrt{\pi}}{(4 \pi f)^2}
\int_0^\infty \frac{ds}{s^{3/2}}
\frac{e B s}{\sinh e B s}
e^{ - ( m_\pi^2 - \Delta^2) s} 
\Erfc
\left( \Delta \sqrt{s} \right)
\left[
\vartheta_3( 0, e^{ - \frac{L^2}{4s}}) - 1
\right]
\label{eq:FVmomentNDeltaFV}
,\end{eqnarray}
from delta-pion contributions to the magnetic moment, 
and
\begin{eqnarray}
\Delta \beta_M^{(\Delta)} (B, L)
&=&
\frac{8}{9}
\frac{g_{\D N}^2 \sqrt{\pi}}{(4 \pi f)^2}
\int_0^\infty \frac{ds}{s^{3/2}}
\left(
\frac{e B s}{\sinh e B s} - 1
\right)
e^{ - m_\pi^2  s} 
\left[
\frac{\Delta}{\sqrt{\pi s}}
+
\left( m_\pi^2 - \Delta^2 \right)
e^{ \Delta^2 s} 
\Erfc
\left( \Delta \sqrt{s} \right)
\right]
\notag \\
&& \phantom{spacingspacingspacing}
\times \left[
\vartheta_3( 0, e^{ - \frac{L^2}{4s}}) - 1
\right]
\Big/ \left( - \frac{1}{2} 4 \pi B^2 \right)
\label{eq:betaFVDeltaFV}
,\end{eqnarray}
for the magnetic polarizability. 
Jacobi elliptic-theta functions appear in this expressions, 
and are defined in Appendix~\ref{s:A}.

\section{Jacobi Elliptic-Theta Functions}
\label{s:A}

Displayed below are one-dimensional, 
proper-time integral representations for the finite volume effects. 
The sums over winding number are cast in terms of three out of the four Jacobi elliptic-theta functions, 
namely
\begin{equation}
\vartheta_2(z,q) 
=
2 
\sum_{n = 0}^\infty
q^{(n + \frac{1}{2})^2} \cos [( 2 n +1) z ]
,
\quad
\vartheta_3(z,q)
=
1 
+ 
2 
\sum_{n = 0}^\infty
q^{n^2} \cos ( 2 n z )
, 
\quad
\vartheta_4(z,q)
=
1 
+ 
2 
\sum_{n = 0}^\infty
(-)^n
q^{n^2} \cos ( 2 n z )
.\end{equation}
As is customary, 
primes denote derivatives with respect to the first argument, 
$\vartheta'_j (z,q) = \frac{d}{dz} \vartheta_j ( z, q)$. 
For both the spin-independent 
$(j = 0)$
and 
spin-dependent
$(j = 1)$
neutron potentials, 
we write
\begin{eqnarray}
V_j(\vec{x}_\perp)
&=& 
\frac{g_A^2 \sqrt{\pi}}{(4 \pi f)^2}
\int_0^\infty \frac{ds}{s^{3/2}}
\frac{e B s}{\sinh e B s}
e^{ - m_\pi^2 s} \,
\mathcal{U}_j (s, B, \vec{x}_\perp)
.\end{eqnarray}
For both the spin-independent and spin-dependent cases, 
we separate out two recurrent terms
\begin{eqnarray}
\mathcal{U}_0 ( s, B, \vec{x}_\perp )
&=&
m_\pi^2 \, 
\mathcal{U}^{(A)} (s , B, \vec{x}_\perp)
+ 
\frac{1}{2} ( e B L)^2 \,
\mathcal{U}^{(B)} (s , B, \vec{x}_\perp),
\\
\mathcal{U}_1 (s, B, \vec{x}_\perp)
&=&
- 
\mathcal{U}^{(A)} (s, B, \vec{x}_\perp)
+ 
\frac{e B L^2}{2 \tanh e B s}
\mathcal{U}^{(B)} (s, B, \vec{x}_\perp) 
\label{eq:spindepU1}
,\end{eqnarray}
with the ancillary functions 
$\mathcal{U}^{(A)} (s, B, \vec{x}_\perp)$
and
$\mathcal{U}^{(B)} (s, B, \vec{x}_\perp)$
having the definitions
\begin{eqnarray}
\mathcal{U}^{(A)} (s, B, \vec{x}_\perp)
=
\begin{cases}
\vartheta_3(\frac{1}{2} e B L x_1 ,e^{- \frac{e B L^2}{4 \tanh e B s}})
\vartheta_3(\frac{1}{2} e B L x_2 ,e^{- \frac{e B L^2}{4 \tanh e B s}})
\vartheta_3(0,e^{- \frac{L^2}{4s}})
-1
, 
&
N_\Phi = \text{even}
\\
\\
\vartheta_3(e B L x_1 ,e^{- \frac{e B L^2}{\tanh e B s}})
\vartheta_3(\frac{1}{2} e B L x_2 ,e^{- \frac{e B L^2}{4 \tanh e B s}})
\vartheta_3(0,e^{- \frac{L^2}{4s}})
-1
\\
\phantom{mm}
+
\vartheta_2(e B L x_1 ,e^{- \frac{e B L^2}{\tanh e B s}})
\vartheta_4(\frac{1}{2} e B L x_2 ,e^{- \frac{e B L^2}{4 \tanh e B s}})
\vartheta_3(0,e^{- \frac{L^2}{4s}})
, 
&
N_\Phi = \text{odd},
\end{cases}
\label{eq:UA}
\end{eqnarray}
along with
\begin{eqnarray}
\mathcal{U}^{(B)} (s, B, \vec{x}_\perp)
=
\begin{cases}
\Big[
- 
\frac{1}{4}
\vartheta''_3(\frac{1}{2} e B L x_1 ,e^{- \frac{e B L^2}{4 \tanh e B s}})
\vartheta_3(\frac{1}{2} e B L x_2 ,e^{- \frac{e B L^2}{4 \tanh e B s}})
\\
\phantom{mm}
- 
\frac{1}{4}
\vartheta_3(\frac{1}{2} e B L x_1 ,e^{- \frac{e B L^2}{4 \tanh e B s}})
\vartheta''_3(\frac{1}{2} e B L x_2 ,e^{- \frac{e B L^2}{4 \tanh e B s}})
\Big]
\vartheta_3(0,e^{- \frac{L^2}{4s}})
, 
&
N_\Phi = \text{even}
\\
\\
\Big[
- 
\vartheta''_3(e B L x_1 ,e^{- \frac{e B L^2}{\tanh e B s}})
\vartheta_3(\frac{1}{2} e B L x_2 ,e^{- \frac{e B L^2}{4 \tanh e B s}})
\\
\phantom{mm}
- 
\frac{1}{4}
\vartheta_3(e B L x_1 ,e^{- \frac{e B L^2}{\tanh e B s}})
\vartheta''_3(\frac{1}{2} e B L x_2 ,e^{- \frac{e B L^2}{4 \tanh e B s}})
\\
\phantom{mm}
- 
\vartheta''_2(e B L x_1 ,e^{- \frac{e B L^2}{\tanh e B s}})
\vartheta_4(\frac{1}{2} e B L x_2 ,e^{- \frac{e B L^2}{4 \tanh e B s}})
\\
\phantom{mm}
- 
\frac{1}{4}
\vartheta_2(e B L x_1 ,e^{- \frac{e B L^2}{\tanh e B s}})
\vartheta''_4(\frac{1}{2} e B L x_2 ,e^{- \frac{e B L^2}{4 \tanh e B s}})
\Big]
\vartheta_3(0,e^{- \frac{L^2}{4s}})
, 
&
N_\Phi = \text{odd}.
\end{cases}
\label{eq:UB}
\end{eqnarray}
As defined, 
these finite volume effects include those for vanishing fields, 
such as the mass renormalization. 
These field-independent contributions can be removed by subtraction
\begin{equation}
\Delta V_j(\vec{x}_\perp)
=
V_j (\vec{x}_\perp) 
- 
\frac{g_A^2 \sqrt{\pi}}{(4 \pi f)^2}
\int_0^\infty \frac{ds}{s^{3/2}}
e^{ - m_\pi^2 s} \,
\mathcal{U}_j (s, 0, 0)
,\end{equation}
where the zero-field functions are considerably simpler;
and, 
are implicitly defined without dependence on the magnetic flux quantum,
$N_\Phi \to 0$. 
Because these finite-volume functions occur frequently, 
we quote them here for ease of reference: 
\begin{equation}
\mathcal{U}_0 (s, 0, 0)
= 
m_\pi^2 \,
\mathcal{U}^{(A)} (s, 0, 0)
, 
\qquad
\mathcal{U}_1 (s, 0, 0)
=
- 
\mathcal{U}^{(A)} (s, 0, 0)
+ 
\frac{L^2}{2 s}
\mathcal{U}^{(B)} (s, 0, 0) 
\label{eq:U0s}
,\end{equation}
with
\begin{eqnarray}
\mathcal{U}^{(A)} (s, 0, 0)
=
\vartheta_3(0,e^{- \frac{L^2}{4s}})^3
-1
, 
\quad
\text{and}
\quad
\mathcal{U}^{(B)} (s, 0, 0)
=
- 
\frac{1}{2}
\vartheta''_3(0, e^{- \frac{L^2}{4 s}})
\vartheta_3(0, e^{- \frac{L^2}{4s}})^2
\label{eq:UAB0s}
.\end{eqnarray}

An alternate way to write the finite volume corrections to the magnetic moment and magnetic polarizability shown in 
Eqs.~\eqref{eq:FVmoment} and \eqref{eq:betaFV}, 
respectively, 
is in terms of the ancillary functions
$\mathcal{U}^{(A)} (s, 0, 0)$
and
$\mathcal{U}^{(B)} (s, 0, 0)$, 
namely
\begin{eqnarray}
\Delta \kappa_n (L)
=
\frac{2 g_A^2 M \sqrt{\pi}}{(4 \pi f)^2}
\int_0^\infty \frac{ds}{s^{3/2}}
e^{ - m_\pi^2 s} \,
\mathcal{U}_1 (s,0,0)
,\end{eqnarray}
for the magnetic moment, 
and
\begin{eqnarray}
\Delta \beta_M (L)
&=&
\frac{e^2}{4 \pi}
\frac{g_A^2 \sqrt{\pi}}{(4 \pi f)^2}
\int_0^\infty \frac{ds}{s^{3/2}}
e^{ - m_\pi^2 s}
\left[
\frac{1}{3} m_\pi^2 s^2  \,
\mathcal{U}^{(A)} (s, 0, 0)
-
L^2
\left( 1 - \frac{1}{6} m_\pi^2 s \right)
\mathcal{U}^{(B)} (s,0,0) 
\right]
,\end{eqnarray}
for the magnetic polarizability. 
Recall that these results only apply in extreme limits.

\section{Equivalence of Finite Volume Corrections to the Magnetic Moment}
\label{s:B}

Here we demonstrate the equivalence of finite volume corrections to the nucleon magnetic moment computed in two different ways. 
The first way employs an external magnetic field, 
and is the result derived above in the limit of an extremely weak magnetic field, 
namely that appearing in Eq.~\eqref{eq:FVmoment}. 
The second way employs isospin twisted boundary conditions in the computation of the nucleon isovector magnetic form factor, 
for which the result can be found in~\cite{Tiburzi:2006px}.
To compare these two methods, 
we take the limit of zero twist angle of the latter computation. 
In order to compare results, 
we must be careful to multiply those obtained in%
~\cite{Tiburzi:2006px}
by a factor of 
$- \frac{1}{2}$
in order to convert loop contributions for the isovector magnetic moment into loop contributions for the neutron magnetic moment. 
In this context, 
it is useful to recall that the one-loop chiral corrections to the nucleon magnetic moment are exactly isovector.

The isovector spin-flip matrix element between nucleon states is sensitive to the Pauli form factor. 
For a relative twist angle of 
$\theta$
between up and down quarks, 
this matrix element is non-vanishing for zero Fourier momentum transfer. 
In finite volume,  
the result
\begin{eqnarray}
F^{v}_2
= 
F_2^v (q^2)
+ 
\Delta F_2^v (L)
,\end{eqnarray}
was obtained in~\cite{Tiburzi:2006px}, 
where the effective momentum transfer 
$q$
arises from the twist angle
$\theta$,
through the simple relation 
$q = \theta / L$.  
In the limit, 
$\theta \to 0$, 
the Pauli form factor reduces to the isovector magnetic moment, 
$F_2^v (0) = \kappa^v$.
The contribution to the matrix element 
$\Delta F_2^v (L)$
arises from finite volume effects, 
and was determined to be
\begin{eqnarray}
\Delta F_2^v (L)
&=&
\frac{3 g_A^2 M}{4 \pi^2 f^2} 
\int_0^1 dx \, 
\cL_{33} \Big( m_\pi P_\pi (x,q^2), x q, 0 \Big)
- 
\frac{g_A^2 M}{4 \pi^2 f^2} \frac{ \cK_2 ( m_\pi, q, 0)}{q}
\label{eq:TWBCs}
.\end{eqnarray}
For simplicity, 
we drop contributions with intermediate-state delta resonances. 
One can also verify that the finite volume effects which include delta resonances and are computed in a background magnetic field agree with the zero twist angle limit of the expressions derived in%
~\cite{Tiburzi:2006px}.  
The kinematic function appearing in the first term in 
Eq.~\eqref{eq:TWBCs} 
is given by
$P_\pi (x, q^2) = [ 1 + x ( 1-x) q^2 / m_\pi^2]^{1/2}$, 
while the finite volume functions
$\cL_{33}$
and
$\cK_2$
were defined to be
\begin{eqnarray}
\cL_{33}
(m, q, 0) 
&=& 
\frac{\sqrt{\pi}}{3}
\int_0^\infty \frac{ds}{s^{3/2}}
e^{- m^2 s}
\Bigg[
\vartheta_3 ( 0, e^{ - \frac{L^2}{4 s}} )^2
\vartheta_3 (\frac{q L}{2}, e^{ - \frac{L^2}{4 s}} )
- 
1
+ 
\frac{L^2}{8 s}
\vartheta''_3 ( 0, e^{ - \frac{L^2}{4 s}} )
\vartheta_3 ( 0, e^{ - \frac{L^2}{4 s}} )
\vartheta_3 (\frac{q L}{2}, e^{ - \frac{L^2}{4 s}} )
\Bigg]
\notag \\
\\
\mathcal{K}_2 (m, q, 0)
&=&
- 
\frac{\sqrt{\pi}L}{4}
\int_0^\infty \frac{ds}{s^{5/2}}
e^{ - m^2 s}
\vartheta'_3 (\frac{q L}{2}, e^{ - \frac{L^2}{4 s}} )
\vartheta_3 ( 0, e^{ - \frac{L^2}{4 s}} )^2
.\end{eqnarray}
The Jacobi elliptic-theta functions appearing in these definitions are given in Appendix~\ref{s:A}.

To take the 
$\theta \to 0$
limit of the finite volume correction 
$\Delta F_2^v (L)$, 
we observe that the first term in Eq.~\eqref{eq:TWBCs} can merely be evaluated at 
$q =0$, 
while the second term requires a derivative, 
$\lim_{q \to 0} \cK_2 ( m, q, 0) / q = \frac{\partial}{\partial q} \cK_2 ( m, 0, 0)$. 
Upon taking the limit, 
we thus find
\begin{eqnarray}
\Delta \kappa^v (L)
\equiv
\lim_{\theta \to 0}
\Delta F_2^v ( L)
= 
-
\frac{g_A^2 \sqrt{\pi} M}{4 \pi^2 f^2}
\int_0^\infty \frac{ds}{s^{3/2}} 
e^{ - m_\pi^2 s} \,
\mathcal{U}_1 ( s, 0 ,0)
=
-
\frac{g_A^2 M m_\pi}{3 \pi f^2}
\sum_{\vec{\nu} - \{\vec{0} \}}
\left[
1
-
\frac{1}{2 |\vec{\nu}| m_\pi  L}
\right]
e^ {- |\vec{\nu}| m_\pi L}
,\end{eqnarray}
where 
$\mathcal{U}_1 (s, 0, 0)$
is the function appearing in the context of finite volume corrections to the magnetic moment, 
and is given in 
Eq.~\eqref{eq:U0s}. 
The second equality arises from performing the proper-time integration. 
Multiplying this finite volume correction to the isovector magnetic moment by 
$- \frac{1}{2}$
leads to the finite volume correction to the neutron magnetic moment, 
which agrees with 
Eq.~\eqref{eq:FVmoment}.

\end{widetext}

\bibliography{magtron}

\begin{thebibliography}{43}
\expandafter\ifx\csname natexlab\endcsname\relax\def\natexlab#1{#1}\fi
\expandafter\ifx\csname bibnamefont\endcsname\relax
  \def\bibnamefont#1{#1}\fi
\expandafter\ifx\csname bibfnamefont\endcsname\relax
  \def\bibfnamefont#1{#1}\fi
\expandafter\ifx\csname citenamefont\endcsname\relax
  \def\citenamefont#1{#1}\fi
\expandafter\ifx\csname url\endcsname\relax
  \def\url#1{\texttt{#1}}\fi
\expandafter\ifx\csname urlprefix\endcsname\relax\def\urlprefix{URL }\fi
\providecommand{\bibinfo}[2]{#2}
\providecommand{\eprint}[2][]{\url{#2}}

\bibitem[{\citenamefont{Martinelli et~al.}(1982)\citenamefont{Martinelli,
  Parisi, Petronzio, and Rapuano}}]{Martinelli:1982cb}
\bibinfo{author}{\bibfnamefont{G.}~\bibnamefont{Martinelli}},
  \bibinfo{author}{\bibfnamefont{G.}~\bibnamefont{Parisi}},
  \bibinfo{author}{\bibfnamefont{R.}~\bibnamefont{Petronzio}},
  \bibnamefont{and} \bibinfo{author}{\bibfnamefont{F.}~\bibnamefont{Rapuano}},
  \bibinfo{journal}{Phys.Lett.} \textbf{\bibinfo{volume}{B116}},
  \bibinfo{pages}{434} (\bibinfo{year}{1982}).

\bibitem[{\citenamefont{Bernard et~al.}(1982)\citenamefont{Bernard, Draper,
  Olynyk, and Rushton}}]{Bernard:1982yu}
\bibinfo{author}{\bibfnamefont{C.~W.} \bibnamefont{Bernard}},
  \bibinfo{author}{\bibfnamefont{T.}~\bibnamefont{Draper}},
  \bibinfo{author}{\bibfnamefont{K.}~\bibnamefont{Olynyk}}, \bibnamefont{and}
  \bibinfo{author}{\bibfnamefont{M.}~\bibnamefont{Rushton}},
  \bibinfo{journal}{Phys.Rev.Lett.} \textbf{\bibinfo{volume}{49}},
  \bibinfo{pages}{1076} (\bibinfo{year}{1982}).

\bibitem[{\citenamefont{Fiebig et~al.}(1989)\citenamefont{Fiebig, Wilcox, and
  Woloshyn}}]{Fiebig:1988en}
\bibinfo{author}{\bibfnamefont{H.}~\bibnamefont{Fiebig}},
  \bibinfo{author}{\bibfnamefont{W.}~\bibnamefont{Wilcox}}, \bibnamefont{and}
  \bibinfo{author}{\bibfnamefont{R.}~\bibnamefont{Woloshyn}},
  \bibinfo{journal}{Nucl.Phys.} \textbf{\bibinfo{volume}{B324}},
  \bibinfo{pages}{47} (\bibinfo{year}{1989}).

\bibitem[{\citenamefont{Lee et~al.}(2005)\citenamefont{Lee, Kelly, Zhou, and
  Wilcox}}]{Lee:2005ds}
\bibinfo{author}{\bibfnamefont{F.}~\bibnamefont{Lee}},
  \bibinfo{author}{\bibfnamefont{R.}~\bibnamefont{Kelly}},
  \bibinfo{author}{\bibfnamefont{L.}~\bibnamefont{Zhou}}, \bibnamefont{and}
  \bibinfo{author}{\bibfnamefont{W.}~\bibnamefont{Wilcox}},
  \bibinfo{journal}{Phys.Lett.} \textbf{\bibinfo{volume}{B627}},
  \bibinfo{pages}{71} (\bibinfo{year}{2005}), \eprint{hep-lat/0509067}.

\bibitem[{\citenamefont{Christensen et~al.}(2005)\citenamefont{Christensen,
  Wilcox, Lee, and Zhou}}]{Christensen:2004ca}
\bibinfo{author}{\bibfnamefont{J.~C.} \bibnamefont{Christensen}},
  \bibinfo{author}{\bibfnamefont{W.}~\bibnamefont{Wilcox}},
  \bibinfo{author}{\bibfnamefont{F.~X.} \bibnamefont{Lee}}, \bibnamefont{and}
  \bibinfo{author}{\bibfnamefont{L.-m.} \bibnamefont{Zhou}},
  \bibinfo{journal}{Phys.Rev.} \textbf{\bibinfo{volume}{D72}},
  \bibinfo{pages}{034503} (\bibinfo{year}{2005}), \eprint{hep-lat/0408024}.

\bibitem[{\citenamefont{Lee et~al.}(2006)\citenamefont{Lee, Zhou, Wilcox, and
  Christensen}}]{Lee:2005dq}
\bibinfo{author}{\bibfnamefont{F.~X.} \bibnamefont{Lee}},
  \bibinfo{author}{\bibfnamefont{L.}~\bibnamefont{Zhou}},
  \bibinfo{author}{\bibfnamefont{W.}~\bibnamefont{Wilcox}}, \bibnamefont{and}
  \bibinfo{author}{\bibfnamefont{J.~C.} \bibnamefont{Christensen}},
  \bibinfo{journal}{Phys.Rev.} \textbf{\bibinfo{volume}{D73}},
  \bibinfo{pages}{034503} (\bibinfo{year}{2006}), \eprint{hep-lat/0509065}.

\bibitem[{\citenamefont{Shintani et~al.}(2007)\citenamefont{Shintani, Aoki,
  Ishizuka, Kanaya, Kikukawa et~al.}}]{Shintani:2006xr}
\bibinfo{author}{\bibfnamefont{E.}~\bibnamefont{Shintani}},
  \bibinfo{author}{\bibfnamefont{S.}~\bibnamefont{Aoki}},
  \bibinfo{author}{\bibfnamefont{N.}~\bibnamefont{Ishizuka}},
  \bibinfo{author}{\bibfnamefont{K.}~\bibnamefont{Kanaya}},
  \bibinfo{author}{\bibfnamefont{Y.}~\bibnamefont{Kikukawa}},
  \bibnamefont{et~al.}, \bibinfo{journal}{Phys.Rev.}
  \textbf{\bibinfo{volume}{D75}}, \bibinfo{pages}{034507}
  (\bibinfo{year}{2007}), \eprint{hep-lat/0611032}.

\bibitem[{\citenamefont{Engelhardt}(2007)}]{Engelhardt:2007ub}
\bibinfo{author}{\bibfnamefont{M.}~\bibnamefont{Engelhardt}}
  (\bibinfo{collaboration}{LHPC Collaboration}), \bibinfo{journal}{Phys.Rev.}
  \textbf{\bibinfo{volume}{D76}}, \bibinfo{pages}{114502}
  (\bibinfo{year}{2007}), \eprint{0706.3919}.

\bibitem[{\citenamefont{Shintani et~al.}(2008)\citenamefont{Shintani, Aoki, and
  Kuramashi}}]{Shintani:2008nt}
\bibinfo{author}{\bibfnamefont{E.}~\bibnamefont{Shintani}},
  \bibinfo{author}{\bibfnamefont{S.}~\bibnamefont{Aoki}}, \bibnamefont{and}
  \bibinfo{author}{\bibfnamefont{Y.}~\bibnamefont{Kuramashi}},
  \bibinfo{journal}{Phys.Rev.} \textbf{\bibinfo{volume}{D78}},
  \bibinfo{pages}{014503} (\bibinfo{year}{2008}), \eprint{0803.0797}.

\bibitem[{\citenamefont{Aubin et~al.}(2009)\citenamefont{Aubin, Orginos,
  Pascalutsa, and Vanderhaeghen}}]{Aubin:2008qp}
\bibinfo{author}{\bibfnamefont{C.}~\bibnamefont{Aubin}},
  \bibinfo{author}{\bibfnamefont{K.}~\bibnamefont{Orginos}},
  \bibinfo{author}{\bibfnamefont{V.}~\bibnamefont{Pascalutsa}},
  \bibnamefont{and}
  \bibinfo{author}{\bibfnamefont{M.}~\bibnamefont{Vanderhaeghen}},
  \bibinfo{journal}{Phys.Rev.} \textbf{\bibinfo{volume}{D79}},
  \bibinfo{pages}{051502} (\bibinfo{year}{2009}), \eprint{0811.2440}.

\bibitem[{\citenamefont{Detmold et~al.}(2009)\citenamefont{Detmold, Tiburzi,
  and Walker-Loud}}]{Detmold:2009dx}
\bibinfo{author}{\bibfnamefont{W.}~\bibnamefont{Detmold}},
  \bibinfo{author}{\bibfnamefont{B.~C.} \bibnamefont{Tiburzi}},
  \bibnamefont{and}
  \bibinfo{author}{\bibfnamefont{A.}~\bibnamefont{Walker-Loud}},
  \bibinfo{journal}{Phys.Rev.} \textbf{\bibinfo{volume}{D79}},
  \bibinfo{pages}{094505} (\bibinfo{year}{2009}), \eprint{0904.1586}.

\bibitem[{\citenamefont{Alexandru and Lee}(2009)}]{Alexandru:2009id}
\bibinfo{author}{\bibfnamefont{A.}~\bibnamefont{Alexandru}} \bibnamefont{and}
  \bibinfo{author}{\bibfnamefont{F.~X.} \bibnamefont{Lee}},
  \bibinfo{journal}{PoS} \textbf{\bibinfo{volume}{LAT2009}},
  \bibinfo{pages}{144} (\bibinfo{year}{2009}), \eprint{0911.2520}.

\bibitem[{\citenamefont{Detmold et~al.}(2010)\citenamefont{Detmold, Tiburzi,
  and Walker-Loud}}]{Detmold:2010ts}
\bibinfo{author}{\bibfnamefont{W.}~\bibnamefont{Detmold}},
  \bibinfo{author}{\bibfnamefont{B.}~\bibnamefont{Tiburzi}}, \bibnamefont{and}
  \bibinfo{author}{\bibfnamefont{A.}~\bibnamefont{Walker-Loud}},
  \bibinfo{journal}{Phys.Rev.} \textbf{\bibinfo{volume}{D81}},
  \bibinfo{pages}{054502} (\bibinfo{year}{2010}), \eprint{1001.1131}.

\bibitem[{\citenamefont{Alexandru and Lee}(2010)}]{Alexandru:2010dx}
\bibinfo{author}{\bibfnamefont{A.}~\bibnamefont{Alexandru}} \bibnamefont{and}
  \bibinfo{author}{\bibfnamefont{F.}~\bibnamefont{Lee}}, \bibinfo{journal}{PoS}
  \textbf{\bibinfo{volume}{LATTICE2010}}, \bibinfo{pages}{131}
  (\bibinfo{year}{2010}), \eprint{1011.6309}.

\bibitem[{\citenamefont{Freeman et~al.}(2012)\citenamefont{Freeman, Alexandru,
  Lee, and Lujan}}]{Freeman:2012cy}
\bibinfo{author}{\bibfnamefont{W.}~\bibnamefont{Freeman}},
  \bibinfo{author}{\bibfnamefont{A.}~\bibnamefont{Alexandru}},
  \bibinfo{author}{\bibfnamefont{F.}~\bibnamefont{Lee}}, \bibnamefont{and}
  \bibinfo{author}{\bibfnamefont{M.}~\bibnamefont{Lujan}},
  \bibinfo{journal}{PoS} \textbf{\bibinfo{volume}{LATTICE2012}},
  \bibinfo{pages}{015} (\bibinfo{year}{2012}), \eprint{1211.5570}.

\bibitem[{\citenamefont{Primer et~al.}(2013)\citenamefont{Primer, Kamleh,
  Leinweber, and Burkardt}}]{Primer:2013pva}
\bibinfo{author}{\bibfnamefont{T.}~\bibnamefont{Primer}},
  \bibinfo{author}{\bibfnamefont{W.}~\bibnamefont{Kamleh}},
  \bibinfo{author}{\bibfnamefont{D.}~\bibnamefont{Leinweber}},
  \bibnamefont{and} \bibinfo{author}{\bibfnamefont{M.}~\bibnamefont{Burkardt}}
  (\bibinfo{year}{2013}), \eprint{1307.1509}.

\bibitem[{\citenamefont{Lujan et~al.}(2013)\citenamefont{Lujan, Alexandru,
  Freeman, and Lee}}]{Lujan:2013qua}
\bibinfo{author}{\bibfnamefont{M.}~\bibnamefont{Lujan}},
  \bibinfo{author}{\bibfnamefont{A.}~\bibnamefont{Alexandru}},
  \bibinfo{author}{\bibfnamefont{W.}~\bibnamefont{Freeman}}, \bibnamefont{and}
  \bibinfo{author}{\bibfnamefont{F.}~\bibnamefont{Lee}} (\bibinfo{year}{2013}),
  \eprint{1310.4837}.

\bibitem[{\citenamefont{Lujan et~al.}(2014)\citenamefont{Lujan, Alexandru,
  Freeman, and Lee}}]{Lujan:2014kia}
\bibinfo{author}{\bibfnamefont{M.}~\bibnamefont{Lujan}},
  \bibinfo{author}{\bibfnamefont{A.}~\bibnamefont{Alexandru}},
  \bibinfo{author}{\bibfnamefont{W.}~\bibnamefont{Freeman}}, \bibnamefont{and}
  \bibinfo{author}{\bibfnamefont{F.}~\bibnamefont{Lee}} (\bibinfo{year}{2014}),
  \eprint{1402.3025}.

\bibitem[{\citenamefont{'t~Hooft}(1979)}]{'tHooft:1979uj}
\bibinfo{author}{\bibfnamefont{G.}~\bibnamefont{'t~Hooft}},
  \bibinfo{journal}{Nucl.Phys.} \textbf{\bibinfo{volume}{B153}},
  \bibinfo{pages}{141} (\bibinfo{year}{1979}).

\bibitem[{\citenamefont{Smit and Vink}(1987)}]{Smit:1986fn}
\bibinfo{author}{\bibfnamefont{J.}~\bibnamefont{Smit}} \bibnamefont{and}
  \bibinfo{author}{\bibfnamefont{J.~C.} \bibnamefont{Vink}},
  \bibinfo{journal}{Nucl.Phys.} \textbf{\bibinfo{volume}{B286}},
  \bibinfo{pages}{485} (\bibinfo{year}{1987}).

\bibitem[{\citenamefont{Damgaard and Heller}(1988)}]{Damgaard:1988hh}
\bibinfo{author}{\bibfnamefont{P.~H.} \bibnamefont{Damgaard}} \bibnamefont{and}
  \bibinfo{author}{\bibfnamefont{U.~M.} \bibnamefont{Heller}},
  \bibinfo{journal}{Nucl.Phys.} \textbf{\bibinfo{volume}{B309}},
  \bibinfo{pages}{625} (\bibinfo{year}{1988}).

\bibitem[{\citenamefont{Tiburzi}(2013)}]{Tiburzi:2013vza}
\bibinfo{author}{\bibfnamefont{B.}~\bibnamefont{Tiburzi}},
  \bibinfo{journal}{Phys. Rev. D 88,} \textbf{\bibinfo{volume}{034027}},
  \bibinfo{pages}{034027} (\bibinfo{year}{2013}), \eprint{1302.6645}.

\bibitem[{\citenamefont{Schwinger}(1951)}]{Schwinger:1951nm}
\bibinfo{author}{\bibfnamefont{J.~S.} \bibnamefont{Schwinger}},
  \bibinfo{journal}{Phys.Rev.} \textbf{\bibinfo{volume}{82}},
  \bibinfo{pages}{664} (\bibinfo{year}{1951}).

\bibitem[{\citenamefont{Hall et~al.}(2013)\citenamefont{Hall, Leinweber, and
  Young}}]{Hall:2013dva}
\bibinfo{author}{\bibfnamefont{J.}~\bibnamefont{Hall}},
  \bibinfo{author}{\bibfnamefont{D.}~\bibnamefont{Leinweber}},
  \bibnamefont{and} \bibinfo{author}{\bibfnamefont{R.}~\bibnamefont{Young}}
  (\bibinfo{year}{2013}), \eprint{1312.5781}.

\bibitem[{\citenamefont{Al-Hashimi and Wiese}(2009)}]{AlHashimi:2008hr}
\bibinfo{author}{\bibfnamefont{M.}~\bibnamefont{Al-Hashimi}} \bibnamefont{and}
  \bibinfo{author}{\bibfnamefont{U.-J.} \bibnamefont{Wiese}},
  \bibinfo{journal}{Annals Phys.} \textbf{\bibinfo{volume}{324}},
  \bibinfo{pages}{343} (\bibinfo{year}{2009}), \eprint{0807.0630}.

\bibitem[{\citenamefont{Tiburzi and Vayl}(2013)}]{Tiburzi:2012ks}
\bibinfo{author}{\bibfnamefont{B.}~\bibnamefont{Tiburzi}} \bibnamefont{and}
  \bibinfo{author}{\bibfnamefont{S.}~\bibnamefont{Vayl}},
  \bibinfo{journal}{Phys.Rev.} \textbf{\bibinfo{volume}{D87}},
  \bibinfo{pages}{054507} (\bibinfo{year}{2013}), \eprint{1210.4464}.

\bibitem[{\citenamefont{Tiburzi}(2008{\natexlab{a}})}]{Tiburzi:2008ma}
\bibinfo{author}{\bibfnamefont{B.~C.} \bibnamefont{Tiburzi}},
  \bibinfo{journal}{Nucl.Phys.} \textbf{\bibinfo{volume}{A814}},
  \bibinfo{pages}{74} (\bibinfo{year}{2008}{\natexlab{a}}), \eprint{0808.3965}.

\bibitem[{\citenamefont{Gasser and Leutwyler}(1988)}]{Gasser:1987zq}
\bibinfo{author}{\bibfnamefont{J.}~\bibnamefont{Gasser}} \bibnamefont{and}
  \bibinfo{author}{\bibfnamefont{H.}~\bibnamefont{Leutwyler}},
  \bibinfo{journal}{Nucl.Phys.} \textbf{\bibinfo{volume}{B307}},
  \bibinfo{pages}{763} (\bibinfo{year}{1988}).

\bibitem[{\citenamefont{Gasser and Leutwyler}(1987)}]{Gasser:1987ah}
\bibinfo{author}{\bibfnamefont{J.}~\bibnamefont{Gasser}} \bibnamefont{and}
  \bibinfo{author}{\bibfnamefont{H.}~\bibnamefont{Leutwyler}},
  \bibinfo{journal}{Phys.Lett.} \textbf{\bibinfo{volume}{B188}},
  \bibinfo{pages}{477} (\bibinfo{year}{1987}).

\bibitem[{\citenamefont{Cohen et~al.}(2007)\citenamefont{Cohen, McGady, and
  Werbos}}]{Cohen:2007bt}
\bibinfo{author}{\bibfnamefont{T.~D.} \bibnamefont{Cohen}},
  \bibinfo{author}{\bibfnamefont{D.~A.} \bibnamefont{McGady}},
  \bibnamefont{and} \bibinfo{author}{\bibfnamefont{E.~S.}
  \bibnamefont{Werbos}}, \bibinfo{journal}{Phys.Rev.}
  \textbf{\bibinfo{volume}{C76}}, \bibinfo{pages}{055201}
  (\bibinfo{year}{2007}), \eprint{0706.3208}.

\bibitem[{\citenamefont{Jenkins and Manohar}(1991)}]{Jenkins:1990jv}
\bibinfo{author}{\bibfnamefont{E.~E.} \bibnamefont{Jenkins}} \bibnamefont{and}
  \bibinfo{author}{\bibfnamefont{A.~V.} \bibnamefont{Manohar}},
  \bibinfo{journal}{Phys.Lett.} \textbf{\bibinfo{volume}{B255}},
  \bibinfo{pages}{558} (\bibinfo{year}{1991}).

\bibitem[{\citenamefont{Jenkins}(1992)}]{Jenkins:1991ts}
\bibinfo{author}{\bibfnamefont{E.~E.} \bibnamefont{Jenkins}},
  \bibinfo{journal}{Nucl.Phys.} \textbf{\bibinfo{volume}{B368}},
  \bibinfo{pages}{190} (\bibinfo{year}{1992}).

\bibitem[{\citenamefont{Bernard et~al.}(1993)\citenamefont{Bernard, Kaiser, and
  Meissner}}]{Bernard:1993nj}
\bibinfo{author}{\bibfnamefont{V.}~\bibnamefont{Bernard}},
  \bibinfo{author}{\bibfnamefont{N.}~\bibnamefont{Kaiser}}, \bibnamefont{and}
  \bibinfo{author}{\bibfnamefont{U.~G.} \bibnamefont{Meissner}},
  \bibinfo{journal}{Z.Phys.} \textbf{\bibinfo{volume}{C60}},
  \bibinfo{pages}{111} (\bibinfo{year}{1993}), \eprint{hep-ph/9303311}.

\bibitem[{\citenamefont{Beane}(2004)}]{Beane:2004tw}
\bibinfo{author}{\bibfnamefont{S.~R.} \bibnamefont{Beane}},
  \bibinfo{journal}{Phys.Rev.} \textbf{\bibinfo{volume}{D70}},
  \bibinfo{pages}{034507} (\bibinfo{year}{2004}), \eprint{hep-lat/0403015}.

\bibitem[{\citenamefont{Bernard et~al.}(1991)\citenamefont{Bernard, Kaiser, and
  Meissner}}]{Bernard:1991rq}
\bibinfo{author}{\bibfnamefont{V.}~\bibnamefont{Bernard}},
  \bibinfo{author}{\bibfnamefont{N.}~\bibnamefont{Kaiser}}, \bibnamefont{and}
  \bibinfo{author}{\bibfnamefont{U.~G.} \bibnamefont{Meissner}},
  \bibinfo{journal}{Phys.Rev.Lett.} \textbf{\bibinfo{volume}{67}},
  \bibinfo{pages}{1515} (\bibinfo{year}{1991}).

\bibitem[{\citenamefont{Jenkins et~al.}(1993)\citenamefont{Jenkins, Luke,
  Manohar, and Savage}}]{Jenkins:1992pi}
\bibinfo{author}{\bibfnamefont{E.~E.} \bibnamefont{Jenkins}},
  \bibinfo{author}{\bibfnamefont{M.~E.} \bibnamefont{Luke}},
  \bibinfo{author}{\bibfnamefont{A.~V.} \bibnamefont{Manohar}},
  \bibnamefont{and} \bibinfo{author}{\bibfnamefont{M.~J.}
  \bibnamefont{Savage}}, \bibinfo{journal}{Phys.Lett.}
  \textbf{\bibinfo{volume}{B302}}, \bibinfo{pages}{482} (\bibinfo{year}{1993}),
  \eprint{hep-ph/9212226}.

\bibitem[{\citenamefont{Meissner and Steininger}(1997)}]{Meissner:1997hn}
\bibinfo{author}{\bibfnamefont{U.-G.} \bibnamefont{Meissner}} \bibnamefont{and}
  \bibinfo{author}{\bibfnamefont{S.}~\bibnamefont{Steininger}},
  \bibinfo{journal}{Nucl.Phys.} \textbf{\bibinfo{volume}{B499}},
  \bibinfo{pages}{349} (\bibinfo{year}{1997}), \eprint{hep-ph/9701260}.

\bibitem[{\citenamefont{Tiburzi}(2008{\natexlab{b}})}]{Tiburzi:2007ep}
\bibinfo{author}{\bibfnamefont{B.~C.} \bibnamefont{Tiburzi}},
  \bibinfo{journal}{Phys.Rev.} \textbf{\bibinfo{volume}{D77}},
  \bibinfo{pages}{014510} (\bibinfo{year}{2008}{\natexlab{b}}),
  \eprint{0710.3577}.

\bibitem[{\citenamefont{Tiburzi}(2006)}]{Tiburzi:2006px}
\bibinfo{author}{\bibfnamefont{B.~C.} \bibnamefont{Tiburzi}},
  \bibinfo{journal}{Phys.Lett.} \textbf{\bibinfo{volume}{B641}},
  \bibinfo{pages}{342} (\bibinfo{year}{2006}), \eprint{hep-lat/0607019}.

\bibitem[{\citenamefont{Tiburzi}(2009)}]{Tiburzi:2008pa}
\bibinfo{author}{\bibfnamefont{B.~C.} \bibnamefont{Tiburzi}},
  \bibinfo{journal}{Phys.Lett.} \textbf{\bibinfo{volume}{B674}},
  \bibinfo{pages}{336} (\bibinfo{year}{2009}), \eprint{0809.1886}.

\bibitem[{\citenamefont{Detmold et~al.}(2006)\citenamefont{Detmold, Tiburzi,
  and Walker-Loud}}]{Detmold:2006vu}
\bibinfo{author}{\bibfnamefont{W.}~\bibnamefont{Detmold}},
  \bibinfo{author}{\bibfnamefont{B.}~\bibnamefont{Tiburzi}}, \bibnamefont{and}
  \bibinfo{author}{\bibfnamefont{A.}~\bibnamefont{Walker-Loud}},
  \bibinfo{journal}{Phys.Rev.} \textbf{\bibinfo{volume}{D73}},
  \bibinfo{pages}{114505} (\bibinfo{year}{2006}), \eprint{hep-lat/0603026}.

\bibitem[{\citenamefont{Hemmert et~al.}(1998)\citenamefont{Hemmert, Holstein,
  and Kambor}}]{Hemmert:1997ye}
\bibinfo{author}{\bibfnamefont{T.~R.} \bibnamefont{Hemmert}},
  \bibinfo{author}{\bibfnamefont{B.~R.} \bibnamefont{Holstein}},
  \bibnamefont{and} \bibinfo{author}{\bibfnamefont{J.}~\bibnamefont{Kambor}},
  \bibinfo{journal}{J.Phys.} \textbf{\bibinfo{volume}{G24}},
  \bibinfo{pages}{1831} (\bibinfo{year}{1998}), \eprint{hep-ph/9712496}.

\bibitem[{\citenamefont{Hemmert et~al.}(1997)\citenamefont{Hemmert, Holstein,
  and Kambor}}]{Hemmert:1996rw}
\bibinfo{author}{\bibfnamefont{T.~R.} \bibnamefont{Hemmert}},
  \bibinfo{author}{\bibfnamefont{B.~R.} \bibnamefont{Holstein}},
  \bibnamefont{and} \bibinfo{author}{\bibfnamefont{J.}~\bibnamefont{Kambor}},
  \bibinfo{journal}{Phys.Rev.} \textbf{\bibinfo{volume}{D55}},
  \bibinfo{pages}{5598} (\bibinfo{year}{1997}), \eprint{hep-ph/9612374}.

\end{thebibliography}

\end{document}